\documentclass[fleqn,usenatbib]{mnras}
\usepackage[T1]{fontenc}
\usepackage{ae,aecompl}
\usepackage{graphicx}	% Including figure files
\usepackage{amsmath}	% Advanced maths commands
\usepackage{amssymb}	% Extra maths symbols
\usepackage{upgreek,bm}

\newcommand{\psf}{D_\phi(\Delta r)}
\newcommand{\tpg}{\frac{\partial \phi}{\partial r}}
\newcommand{\tpgil}{\partial \phi/\partial r}

\title[Radio wave scattering in the CGM]{Radio wave scattering by circumgalactic cool gas clumps}
\author[Vedantham \& Phinney]{H.~K.~Vedantham$^{1,2}$\thanks{E-mail: vedantham@astron.nl} and E.~S.~Phinney$^{1}$\\
$^{1}$Cahill center for astronomy and astrophysics, California Institute of Technology, 1200 E. California Blvd. Pasadena, CA 91125, USA.\\
$^{2}$Netherlands Institute for Radio Astronomy (ASTRON), Oude Hogeveensedijk 4, 7991 PD, Dwingeloo, Netherlands} 

\date{Accepted XXX. Received YYY; in original form 6 Mar 2018}
\pubyear{2018}

\hypersetup{draft}
\begin{document}
\label{firstpage}
\pagerange{\pageref{firstpage}--\pageref{lastpage}}
\maketitle

\begin{abstract}
We consider the effects of radio-wave scattering by cool ionized clumps ($T\sim 10^4$\,K) in circumgalactic media (CGM). The existence of such clumps are inferred from intervening quasar absorption systems, but have long been something of a theoretical mystery. We consider the implications for compact radio sources of the `fog-like' two-phase model of the circumgalactic medium recently proposed by \citet{mccourt2016}. In this model, the CGM consists of a diffuse coronal gas ($T\gtrsim 10^6$\,K) in pressure equilibrium with numerous $\lesssim 1$\,pc scale cool clumps or `cloudlets' formed by shattering in a cooling instability. The areal filling factor of the cloudlets is expected to exceed unity in $\gtrsim 10^{11.5}\,M_\odot$ haloes, and the ensuing radio-wave scattering is akin to that caused by turbulence in the Galactic warm ionized medium (WIM). If $30$\,per-cent of cosmic baryons are in the CGM, we show that for a cool-gas volume fraction of $f_{\rm v}\sim 10^{-3}$, sources at $z_{\rm s}\sim 1$ suffer angular broadening by $\sim 15\,\upmu$as and temporal broadening by $\sim 1\,$ms at $\lambda = 30$\,cm, due to scattering by the clumps in intervening CGM. The former prediction will be difficult to test (the angular broadening will suppress Galactic scintillation only for $<10\,\upmu$Jy compact synchrotron sources). However the latter prediction, of temporal broadening of localized fast radio bursts, can constrain the size and mass fraction of cool ionized gas clumps as function of halo mass and redshift, and thus provides a test of the model proposed by \citet{mccourt2016}.
\end{abstract}

\begin{keywords}
galaxies: haloes -- scattering -- radio continuum: transients
\end{keywords}

\section{Introduction}
The circumgalactic medium (CGM) of galaxies and the intergalactic medium (IGM) are together expected to harbour about $80$\,\% all baryons in the Universe \citep{anderson2010}. Absorption spectroscopy of quasars along intervening CGM sight-lines in recent years have yielded a wealth of information on the physical state of CGM gas. Some of these findings have however contradicted naive models based on theoretical considerations. In particular, the ubiquitous detection of cool ($\sim 10^4$\,K) and likely dense \citep[$n_e\sim 1$\,cm$^{-3}$ at $z\approx 2$;][]{hennawi2015,lau2016} gas in the CGM of massive galaxies ($M\gtrsim 10^{12}\,M_\odot$) is puzzling--- an outcome that was not predicted by canonical galaxy assembly models. Based on theoretical consideration and numerical simulations, \citet{mccourt2016,ji2017} have shown that numerous sub-parsec scale cool\footnote{The clumps of interest are photoionized gas at $\sim 10^4\mbox{K}$. In the recent CGM literature, whose terminology is used by  McCourt et al, such clouds are called ``cold'', while gas at $10^{4-5}\mbox{K}$ is called  ``cool'', and gas at $10^{5-6}\mbox{K}$ called ``warm'', in contrast to the volume-filling ``hot'' gas at $\sim 10^6\mbox{K}$. This is unfortunately inconsistent with many decades of tradition of literature on the interstellar medium, in which partially ionized gas at $8000\mbox{K}$ is called ``warm'', while ``cold'' is reserved for neutral and molecular gas at much lower temperatures. To avoid confusion, in this paper we decided to call the photoionized clumps ``cool'' rather than ``cold'' or ``warm''.} gas clumps can form in galaxy haloes due to thermal instabilities, likening the CGM to a `fog' consisting of partially ionized $\sim 10^4$\,K cloudlets dispersed in a hot $\sim 10^6$\,K ambient medium. Such small clumps, though they can explain many features of quasar absorption lines, are however subject to uncertainly regarding the initial conditions, and destruction by electron conduction from the surrounding hot gas unless magnetically shielded.  It therefore is desirable to have an observational probe capable of detecting the existence of such small clumps in the CGM of distant galaxies. Here we show that the fog-like CGM leads to observable scattering of radio waves from extragalactic sources, and that upcoming surveys for Fast Radio Bursts (FRB) can constrain the sub-parsec scale morphology of cool gas in intervening CGM.

The assembly of dissipative baryons into galaxies in the presence of dark matter potential wells has been studied extensively based on general physical principles\citep[see e.g.][]{binney1977, silk1977, rees1977, white1978}. The conclusion of these early studies relevant for this paper is as follows. Gravitational potential energy of baryons is converted to kinetic energy during dissipative collapse. This heats the baryons to the virial temperature $T_{\rm vir}\approx 10^6\,M_{12}^{2/3}h^{2/3}(z)$\,K, where $M_{12}$ is the halo mass in units of $10^{12}\,M_\odot$ and $h(z)$ is the dimensionless Hubble parameter at redshift $z$ of the halo. For haloes less massive than about $10^{11.5}\,M_\odot$, the virial temperature drops below $\sim 10^{5}\,$K, whereupon the radiative cooling timescale, $t_{\rm cool}$ (primarily via metal lines) is smaller than the Hubble time, $t_0$. The gas then rapidly cools to $T\sim 10^4\,$K throughout the halo, loses pressure support, and falls inwards to form stars. In more massive haloes, $t_{\rm cool}\gg t_0$ and the gas at large radii $r\sim r_{\rm vir}$ forms a long-lived pressure-supported hot ($T\gtrsim 10^{5.5}\,$K) halo devoid of cold neutral gas.\footnote{At some sufficiently small radius $r\ll r_{\rm vir}$, $t_{\rm cool}<t_0$ and the gas can collapse into stars.} 

Quasar absorption spectroscopy and fluorescent Ly$\upalpha$ studies of the CGM, however, tell a somewhat different story. Absorption studies routinely detected large amounts ($N\sim 10^{18}-10^{20}$\,cm$^{-2}$) of cool $10^4$\,K gas at the virial radius of $\gtrsim 10^{12}\,M_\odot$ haloes at both high ($z\sim 2$) and low ($z\sim 0$) redshifts \citep{steidel1988,steidel1994,lau2016,stocke2013,werk2014,mathes2017,tumlinson2017}.  Studies of galaxies at redshifts $z\sim 0.1-2.5$ have shown that associated absorption lines are almost always found in the spectra of background quasars projected within 50--100~kpc of the galaxies\citep{tumlinson2017,rudie2012,turner2014}. Thus the covering factor of cool gas in galaxy halos exceeds 50\%\ even at such large distances from the galaxies. Photoionization models, though uncertain, indicate that the projected mass surface density scales roughly as $r^{-1}$ \citep[][see e.g., Fig 7 of]{tumlinson2017}. Florescent Ly$\upalpha$ imaging of quasar host galaxies at $z\sim 3$ provides additional cofirmation that cool gas has a covering fraction of unity even out to the virial radius, with a surface brightness that evolves with radius as $r^{-1.8}$ \citep{cantalupo2014,borisova2016}. Both the emission and absorption observations point to the ubiquitous nature of cool, $\sim 10^4$\,K gas in the CGM of $M\gtrsim 10^{12}\,M_\odot$ halos--- a result not predicted by the canonical model of halo formation. The radial profile and smooth absorption lines over the viral velocity width in addition disfavour any model where the cool gas is confined to a narrow shell around the virial shock, but instead suggests that the cool gas pervades the CGM in multiple small clouds with a total areal covering factor exceeding unity. 

Broadly speaking, two classes of models have been advanced via sophisticated simulations to explain the large covering fraction of cool gas in massive galactic haloes. (a) The first set of models create the cool gas in-situ by recognizing that in practice, only a part of the accreted gas is likely heated to the virial temperature at the accretion shock \citep[see e.g.][fig. 7]{keres2005} and/or by enhancing thermal instability via magnetic suppression of buoyant oscillation \citep{ji2017}. The cooler ($T\lesssim 10^{5.5}$) gas can therefore cool well within $t_0$ {\em in situ}. (b) The second set of models transport the cool gas from near the galactic disk into the halo in the form of galactic winds \citep{fg2016}. At present, these are somewhat heuristic arguments and the precise details of how cool gas is produced and sustained in galactic haloes remains an active field of study (see \citet{tumlinson2017} for a recent review).

Absorption spectroscopy \citep[][e.g.]{steidel1988,tumlinson2017} measures the column density of cool gas. The volume density can only be inferred from photoionization modelling that is fraught with uncertainties. Yet the volume density of cool gas in CGM and its internal clumpy structure are critical to a determination of its physical state, formation mechanism, and eventual fate. Recently \citet{mccourt2016} have employed simulations and theoretical arguments to study the condensation of cool $T\approx 10^4$\,K clumps from a background of hot $T\gtrsim 10^6\,$K gas that leads to the development of the classical two-phase medium \citep{field1965}. They argue that, akin to fragmentation in the Jeans' instability \citep{jeans1902} of gravitationally collapsing clouds with $\gamma<4/3$, cooling of clouds whose size greatly exceeds $c_s t_{\rm cool}$ does not proceed isochorically, but leads to continual fragmentation of gas into pieces of size $\sim c_s t_{\rm cool}$ which are able to maintain isobaric cooling, down to a length-scale of order the minimum of $c_s(T)t_{\rm cool}(T)$ as a function of temperature $T$. For radiative cooling curves relevant to astrophysical plasma, this characteristic minimum scale of cool clumps occurs at $T\sim 10^4$ and is $\sim (0.1\,{\rm pc})\,(n/{\rm cm}^{-3})^{-1}$. This predicts a fixed gas column density of the individual smallest clumps (independent of ambient pressure) of $N_e\approx 10^{17}\,$cm$^{-2}$ \citep[][their \S 2.1]{mccourt2016}. Such small length-scales are currently well beyond the reach of halo-scale simulations and much smaller than can be constrained by photoionization modelling of absorption spectra. 

By contrast, the scattering of radio waves is a highly sensitive function of small-scale density inhomogeneities. For instance, radio wave propagation through the Galactic warm ionized medium (WIM) has been used to study its density structure on spatial scales of $10^8-10^{15}\,$cm \citep{armstrong1995}. Here we show that the same techniques can be applied to probe the structure of cool gas in the CGM. More importantly, the recent discovery of Fast Radio Bursts \citep[FRB;][]{lorimer2007}--- milli-second duration radio pulses originating at cosmic distances, opens up an unprecedented opportunity to revolutionize our understanding of the CGM, much in the same way the discovery of pulsars led to a profoundly improved understanding of the Galactic interstellar medium \citep{rickett1990}.

The rest of the paper is organized as follows. In \S 2, we lay down the basic halo properties as as function of mass and redshift. In \S 3 we compute the scattering characteristic of such haloes. In \S 4, we present a discussion of our results by considering the observable signature of scattering by the CGM of an ensemble of haloes in the Universe.
We adopt the {\em Planck} cosmological parameters \citep{plank2015}: $H_0=67.8\,$km\,s$^{-1}$\,Mpc$^{-1}$, $\Omega_{\rm m} = 0.308$ and $\Omega_\Lambda = 1-\Omega_{\rm m}$ throughout this paper. A glossary of symbols and their meaning is given in the Appendix for quick reference.

\section{Halo properties}
\label{sec:halo_properties}
We assume the usual definition of virial radius, $r_{200}$ as the radius at which the matter density equals 200 times the critical density at any given redshift. The halo mass, $M_{12}=M/10^{12}\,M\odot$ is then the mass enclosed within $r_{200}$:
\begin{equation}
\label{eqn:r200}
r_{200} = \left( \frac{3M}{800\uppi \rho(z)}\right)^{1/3} \approx 163\,M_{12}^{1/3}\,h^{-2/3}(z)\,{\rm kpc,}
\end{equation}
where the critical density $\rho(z)$ is given by
\begin{equation}
\label{eqn:rhoz}
\rho(z) = \frac{3H(z)^2}{8\uppi G}\approx 277.34\, h^2(z)\,\,M_{\sun}\,{\rm kpc}^{-3}.
\end{equation}
Here $H(z)$ is the Hubble parameter at redshift $z$:
\begin{equation}
\label{eqn:Hz}
H(z) = H_0\sqrt{\Omega_{\rm m}(1+z)^3 + \Omega_\Lambda},
\end{equation}
and $h(z)$ is the dimensionless Hubble parameter defined as $H(z) = h(z) \times 100\,$km\,s$^{-1}$\,Mpc$^{-1}$. With the dark matter halo properties completely specified by equations \ref{eqn:r200} through to \ref{eqn:Hz}, we turn our attention to the gas properties.

\subsection{Gas density}
We assume that in the halo mass-range of interest, the infalling gas is shock heated to the virial temperature of
\begin{equation}
\label{eqn:tvir}
T_{\rm vir} = 9.3\times10^5\,M_{12}^{2/3}h^{2/3}(z)\,\,{\rm K},
\end{equation}
The hot gas pressure at $r_{200}$ and its profile is somewhat difficult to derive from first principles. We therefore pick a gas pressure at $r_{200}$ that yields a predefined baryon fraction in the hot phase. There is currently no consensus on how the cosmic baryons are apportioned to the various gas and stellar phases. Current best constraints places about $20\,$per-cent of baryons in galaxies \citep[stars, gas and dust; see e.g.][]{anderson2010}, while the remaining $80$\,per-cent must be in the CGM and IGM \citep{tumlinson2017}. We will normalize our results to the nominal case where $f_{\rm CGM} = 30$\,per-cent of baryons are in the CGM. We further assume a density profile of $n(r)\propto r^{-\alpha}$ for $0<r\le r_{\rm shock}=1.5r_{200}$ with $\alpha=1.5$ \citep{fielding2017}. This yields a gas pressure at $r_{200}$ of
\begin{equation}
\label{eqn:p200}
P_{200}(M,z) = 27\,\left(\frac{f_{\rm CGM}}{0.3}\right)\,M_{12}^{2/3}\,h(z)^{8/3}\,\,{\rm cm}^{-3}\,{\rm K}.
\end{equation}

\subsection{Volume fraction and covering factor}
\begin{figure*}
\centering
\includegraphics[width=0.65\linewidth]{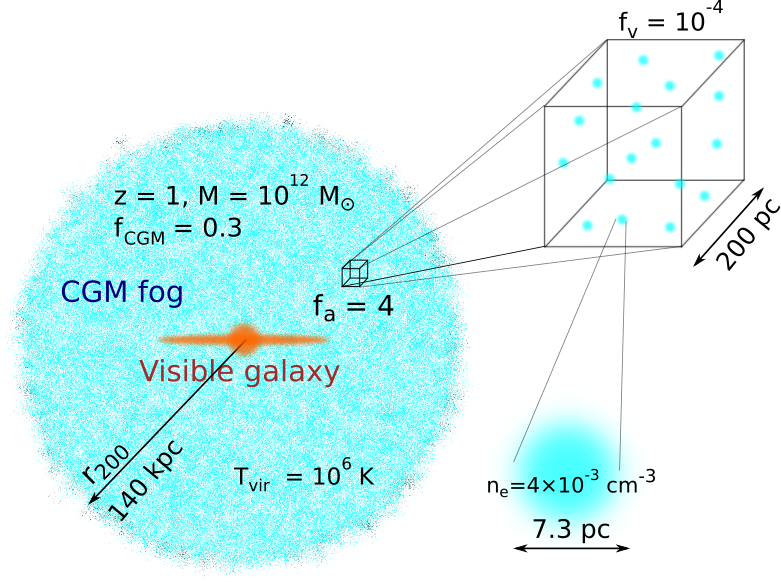}
\caption{A depiction of the fog-like CGM model considered here and some characteristics physical parameters for a $10^{12}\,M_\odot$ halo at $z=1$. Cyan denotes the cool $10^4$\,K gas clumps or `cloudlets' that are dispersed in a virial-shock heated $10^6$\,K halo gas. The cloudlets have a large areal covering factor despite their small volume fraction. \label{fig:fog_sketch}}
\end{figure*}
\begin{figure*}
\centering
\includegraphics[width=\linewidth]{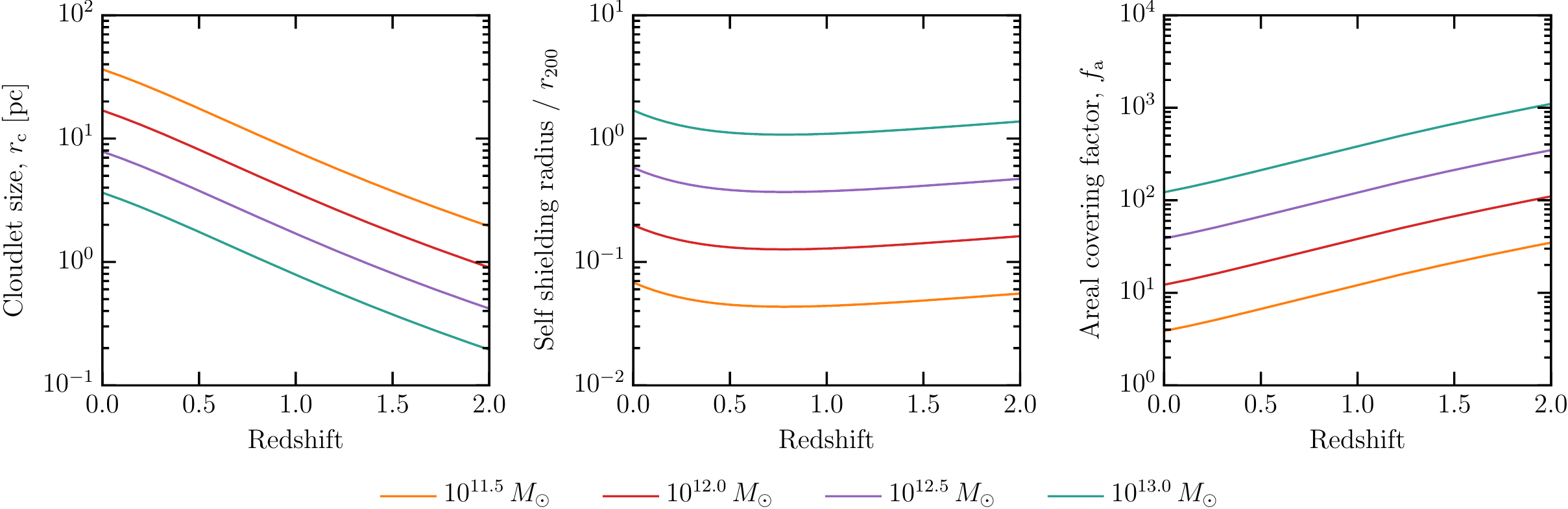}
\caption{CGM fog characteristics assumed in this paper (equations \ref{eqn:r200} to \ref{eqn:fa}). The column density through individual cloudlets and the cloudlet volume fraction are taken to be  $N_e=10^{17}\,$cm$^{-2}$, and $f_{\rm v}=10^{-4}$ respectively. The cloudlet size, $r_{\rm c}$ and the areal covering factor, $f_{\rm a}$ are evaluated at the viral radius, $r_{\rm 200}$, and evolve with radius according $r^{1.5}$ and $r^{0.2}$ respectively \label{fig:rc_rss_fa}}
\end{figure*}
The closest analog to the cool clumps that have been studied in any detail are the Milky Way's high velocity clouds (HVC). HVCs detected in emission have a total mass of about $2.6\times 10^7\,M_\odot$ \citep{putman2012} which yields a lower limit on cool gas volume fraction of about $f_{\rm v}>10^{-5}$. Several authors have studies absorption line systems at higher redshifts up to $z\sim 2$. The measured column densities in conjunction with photo-ionization modeling yield volume fractions of $f_{\rm v}=10^{-4}-10^{-3.5}$ \citep{lau2016,hennawi2015,prochaska2009,stocke2013}. We refer the reader to \citet[][their Table 1]{mccourt2016} for a summary of these results. 
Photo-ionization modeling sufferers from considerable uncertainties. In limited cases, fine structure lines may be used to get a direct estimate of gas densities without the need for photo-ionization modeling. Such observations also show large gas densities\cite{lau2016} in excess of $\sim 1\,{\rm cm}^{-3}$ that imply comparable volume filling fractions. Hence, we will adopt a characteristic of $f_{\rm v}=10^{-4}$ when a specific number is required, but we will carry $f_{\rm v}$ as a variable in our equations such that variations between photo-ionization models may be included in the future.

There is sparse observational constraint regarding the radial evolution of the volume fraction. \citet{borisova2016} find that the surface brightness of the Ly$\upalpha$ fluorescent emission in $z\sim 2-3$ CGM has a power-law variation, $r^{-1.8}$. If the volume density of cool gas evolves as $r^{-1.5}$ as seen in simulations \citep{fielding2017}, then the fluorescent  emission can be reconciled with a volume fraction that changes only weakly with radius as $f_{\rm v}(r)\propto r^{-\beta}$, with $\beta=-0.2$. We will adopt this value throughout.

The foggy-CGM model under consideration here specifically addresses the large areal covering factor of cool gas despite its low volume fraction  as implied by photo-ionization modeling. The number of cool clumps encountered by a sightline at a characteristic impact factor of $b$ is given by $f_{\rm a}(b) \approx f_{\rm v}(b)\,b/r_{\rm c}$ where $r_{\rm c}$ is the radius of the individual cloudlets that comprise the CGM fog. Because the column density in individual cloudlets is fixed by the model under consideration at $N_e=10^{17}$\,cm$^{-2}$, the cloudlet size at radius $r$ is $r_{\rm c}(r) = 0.5N_e/n_e(r)$ which gives
\begin{equation}
\label{eqn:fa}
f_{\rm a}(b) \approx 3\,\left(\frac{f_{\rm CGM}}{0.3}\right)\left(\frac{f_{\rm v}}{10^{-4}}\right)\,M_{12}h^2(z)\,\,\left(\frac{b}{r_{200}} \right)^{1-\beta-\alpha}
\end{equation}
For our fiducial values of $\alpha=1.5$, $\beta=-0.2$, $f_{\rm v}=10^{-4}$, $f_{\rm CGM}=0.3$, the covering fraction of cloudlets exceeds unity at impact parameter $b\sim r_{200}$ for haloes above $10^{11.9}\,M_\odot$ at $z=0$ or above $10^{11.4}\,M_\odot$ at $z=1$. This fog-like nature of the CGM wherein a low volume fraction leads to a high areal covering factor is depicted in Fig. \ref{fig:fog_sketch} with some characteristic parameter values for a $10^{12}\,M_\odot$ halo at $z=1$. The reader may readily scale the numbers to other halo masses and redshifts via equations \ref{eqn:r200} though \ref{eqn:fa}. The resulting densities and high areal covering factors are roughly consistent with inferences from observations of quasar absorption systems: see for instance \citet[][their Table. 1]{mccourt2016}, who present a compilation of relevant observational inferences from \citep{stocke2013,lau2016,prochaska2009,hennawi2015}.

\subsection{Neutral fraction}
The last aspect of haloes that needs specification is the ionization fraction, since only free electrons contribute to radio wave scattering. We adopt an intergalactic UV photoionization rate to be $\Gamma({\rm IGM}) = 10^{-13}\left[(1+z)/1.2\right]^{5}$\,sec$^{-1}$ \citep{gaikwad2017} to determine the neutral fraction at different radii, $\zeta(r)$. Details of our photoionization-equilibrium calculations in a fog-like CGM is is given in the Appendix. We find that individual cloudlets with their column density of $10^{17}\,$cm$^{-2}$ are only partially ionized, but the fog can self-shield itself against the extragalactic radiation field below a critical radius that, for $\alpha=1.5,\,\beta=-0.2$, has an approximate value of (proof in Appendix)
\begin{equation}
\label{eqn:zeta_approx}
r_{\rm ss} \approx 0.11\,\left(\frac{f_{\rm v}}{10^{-4}}\right)^{0.56}\,\left(\frac{f_{\rm CGM}}{0.3}\right)^{1.11}M_{12}^{0.93}\,\frac{h^{2.59}(z)}{(1+z)^{2.78}}.
\end{equation}
At radii below $r_{\rm ss}$ the clouds rapidly achieve neutrality. We therefore find that haloes more massive that $10^{13.2}\,M_\odot$ at $z=0$ or $10^{13.4}\,M_\odot$ at $z=1$ can self shield themselves even at their virial radius. The halo mass range that is relevant for radio wave scattering is therefore bounded. On the lower mass end, haloes less massive than $10^{11.5}\,M_\odot$ are not expected to have long-lived pressure supported halo gas, and haloes more massive than about $10^{13.5}\,M_\odot$ can self shield themselves against ionization from the extragalactic radiation field. Now that the halo mass range and the relevant gas properties have been specified, we turn our attention to the problem of computing the scattering parameters.

\section{Scattering by a single halo}
\label{sec:1halo}
Before considering the scattering of radio waves from a cosmic distribution of haloes, it is instructive to built up our analysis starting with the scattering properties of a single cool gas clump, which we will call as a `cloudlet' after \citet{mccourt2016}.
\subsection{Dispersion in a cloudlet}
\begin{figure*}
\centering
\includegraphics[width=0.65\linewidth]{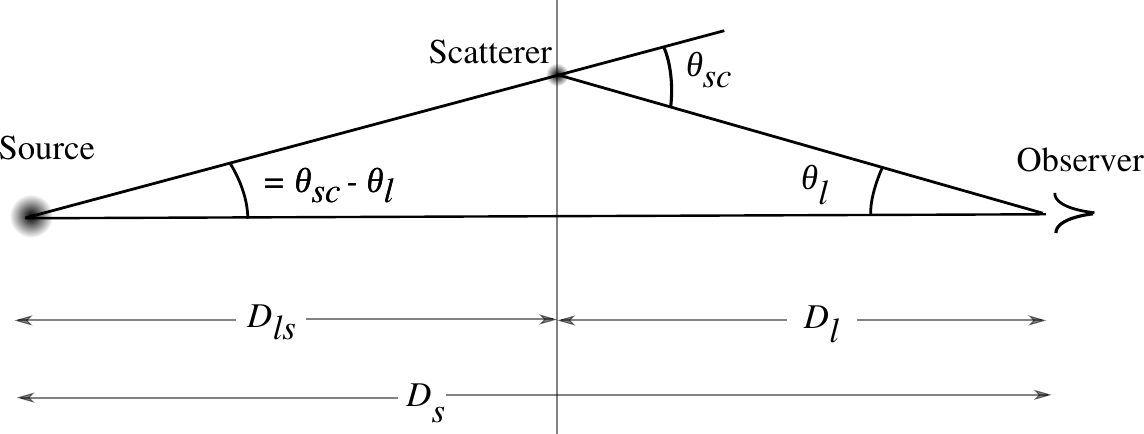}
\caption{A not to scale sketch of the scattering geometry and symbols used in this paper.\label{fig:geom}}
\end{figure*}

Propagation through plasma of column density $N_e$ advances the phase of a monochromatic wave of wavelength $\lambda$ by $\phi = \lambda N_e r_e$ where $r_e$ is the classical electron radius. Wave diffraction is a result of fluctuations of phase $\phi$ transverse to the direction of propagation. Specifically, a transverse gradient of $\tpgil$ leads to a deflection of the direction of light propagation through an angle (geometry sketched in Fig. \ref{fig:geom})
\begin{equation}
\label{eqn:theta_sc}
\theta_{\rm sc} = \frac{\lambda }{2\pi} \tpg 
\end{equation}
For a cloudlet radius of $r_c$, the phase gradient is $\tpgil \sim \lambda N_e r_e/r_{\rm c}$, which gives a characteristic deflection angle of
\begin{equation}
\label{eqn:theta_sc_num}
\theta_{\rm sc} \sim 0.3\,\upmu{\rm as}\,\, \lambda_{30}^2\,  N_{e,17}\, (r_{\rm c}/1\,{\rm pc})^{-1} 
\end{equation}

The geometric delay between the time of arrival of signals from multiple images may also be observed in impulsive sources such as fast radio bursts. Here again, the cosmological distances will have profound effect. The characteristic time delay is
\begin{equation}
\label{eqn:tau}
\Delta \tau \approx \frac{\theta^2_{\rm sc}}{2c}\frac{D_{ls}D_l}{D_s}
\end{equation}
The delay is maximized for a geometry where $D_{ls}=D_l$:
\begin{equation}
\Delta \tau_{\rm max} \approx 0.05\,\upmu{\rm s}\,\,\left(D_l/1\,{\rm Gpc}\right)\, \lambda_{30}^4\, N_{e,17}^2\, \left(r_{\rm c}/1\,{\rm pc}\right)^{-2}
\end{equation}
which is comparable to the temporal broadening due to scattering in the Galactic WIM towards Pulsars at high Galactic latitude \citep[see for e.g.][]{psrcat}. We therefore conclude that even an isolated cloudlet at cosmological distances leads to measurable effects on radio waves. Because we expect typical sight-lines through the CGM to intercept a large number of cloudlets (equation \ref{eqn:fa}), we now generalize these results to the case of a random cloudlet ensemble.

\subsection{Ensemble scattering properties}
\label{subsec:ensemble}
The volume filling fraction of cloudlets is expected to be small: $\sim 10^{-4}$, and we treat the cloudlets as discrete objects that are randomly distributed. The transverse phase gradient imparted by such an ensemble of cloudlets is a random variable, whose statistical properties are best expressed in terms of the phase structure function, $\psf$ defined as:
\begin{equation}
\psf \equiv \left< \left[  \phi(r)-\phi(r+\Delta r) \right]^2\right>
\end{equation}
where the angular brackets denote ensemble average, and $\phi(r)$ is the total wave phase at transverse co-ordinate $r$. The structure function therefore measures the variance of phase differences between two sight-lines that separated by a transverse distance $\Delta r$, and is the statistical analogue of the transverse phase gradient $\tpgil$ used in \S 2.2 to compute the scattering angle. 

If there be on average $f_{\rm a}$ cloudlets intercepted by a sight-line, the structure function can be shown to be (see Appendix)
\begin{equation}
\label{eqn:psf_long}
\psf = 2\lambda^2r_e^2N_e^2f_{\rm a}\Psi(\Delta r/r_c),
\end{equation}
where the function $\Psi(.)\leq 1$ only depends on the internal structure of the cloudlets and determines the slope of the phase structure function. 

\begin{figure}
\centering
\includegraphics[width=\linewidth]{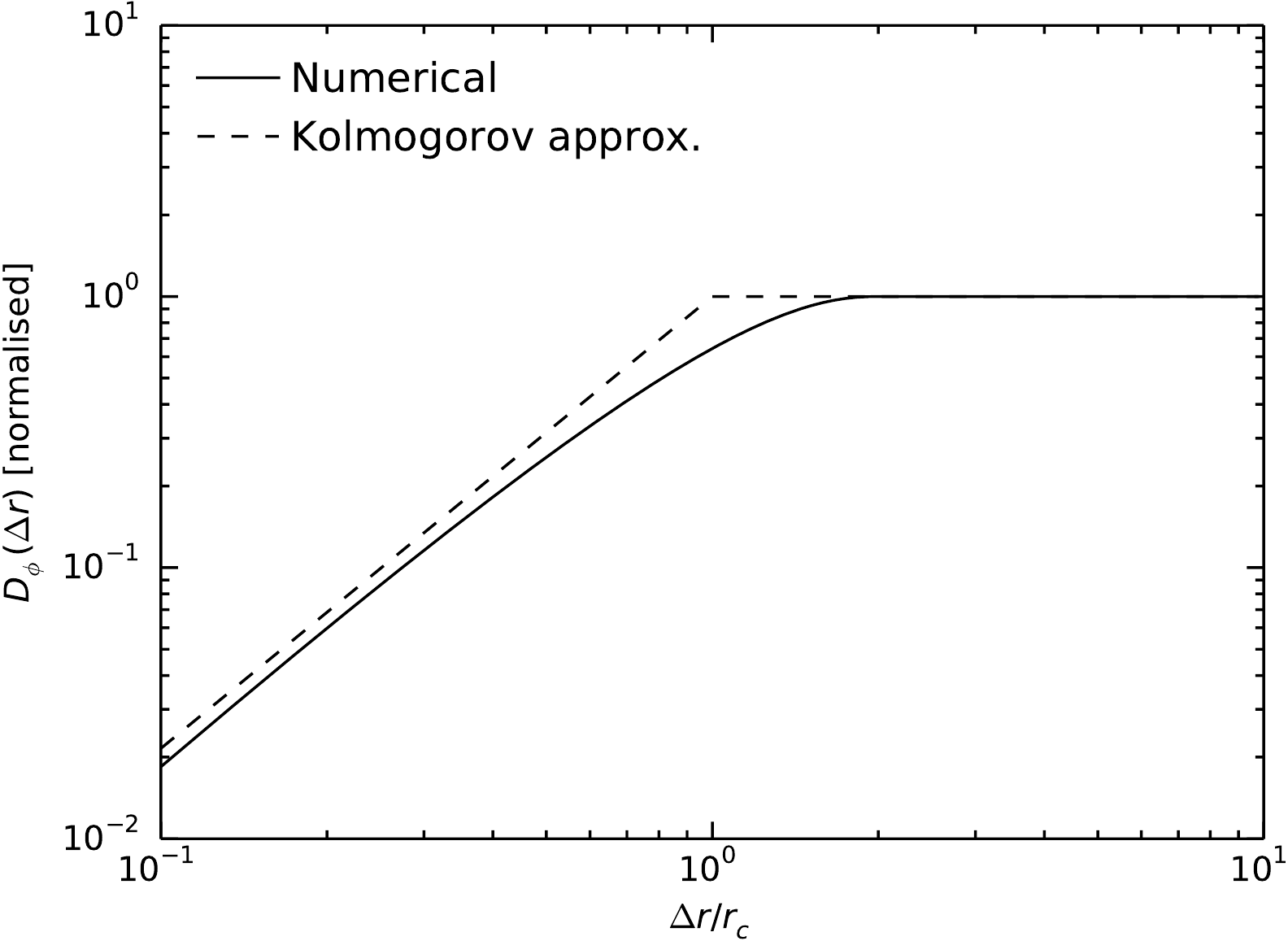}
\caption{The normalize phase structure function of an ensemble of spherical cloudlets with uniform density, compared to a Kolmogorov structure function with the same normalization and an outer-scale of $r_{\rm c}$. The structure function is normalized by the factor $2f_{\rm a}\left(\lambda r_eN_e \right)^2.$\label{fig:psf}}
\end{figure}

The upper panel of Fig. \ref{fig:psf} shows the numerically computed, normalized  structure function $\Psi(\Delta r/r_{\rm c})$ for a spherical cloudlet. As anticipated, the differential phase increases monotonically for $\Delta r<r_c$. Beyond a transverse separation of $r_{\rm c}$, rays encounter an independent realization of cloudlets and the structure function saturates and becomes independent of $\Delta r$. The saturated value of the structure function is simply the Poisson variance in the phase accumulated along two independent realization of the cloudlet ensemble which is $2f_{\rm a}\times (\lambda r_e N_e)^2$. Here the first term is the Poisson variance in the differential number of cloudlets on two independent sight-lines and the second terms if the square of the radio-wave phase through a single cloudlet. 

The bottom panel of Fig. \ref{fig:psf} shows the logarithmic slope of the structure function under the spherical cloudlet assumption. The slope is less than the critical value of $2$, which implies a `shallow spectrum' in which the transverse phase structure on smaller spatial scales dominate the scattering as opposed to larger scales fluctuations \citep{goodman1985}. At smaller spatial scales ($\Delta r \ll r_{\rm c}$), the slope is close to the Kolmogorov value of $5/3$, which is usually employed to model wave scattering in extended turbulent media. We will therefore proceed with the assumption that the structure function is `Kolmogorov-like' with an outer scale of $r_{\rm c}$ and total phase variance of $2\lambda^2N_e^2r_e^2f_{\rm a}$:
\begin{eqnarray}
\label{eqn:psf1}
D_\phi(\Delta r) &=& \left(\frac{\Delta r}{r_{\rm diff}} \right)^{5/3}\,\,\Delta r<r_{\rm c}\nonumber \\
&& 2\lambda^2  N_e^2 r_e^2 f_{\rm a}\,\,{\rm otherwise}
\end{eqnarray}
where, the diffractive scale $r_{\rm diff}$ is 
\begin{equation}
\label{eqn:psf2}
r_{\rm diff} = r_{\rm c}\left(2\lambda^2  N_e^2 r_e^2 f_{\rm a} \right)^{-3/5},
\end{equation}
or
\begin{equation}
\label{eqn:rdiff}
r_{\rm diff} \sim 1.6\times 10^{10}\,{\rm cm}\, \left(r_{\rm c}/1\,{\rm pc} \right) \lambda_{30}^{-6/5}\,N_{e,17}^{-6/5}\left(f_{\rm a}/10\right)^{-3/5}
\end{equation}

For comparison, the Galactic WIM has a diffractive scale of $\sim 10^{9.5}$\,cm at $\lambda = 30$\,cm \citep[][their fig. 2]{armstrong1995}. Hence we expect the cloudlet ensemble to scatter incoming light through angles that are comparable to that from the Galactic WIM.

The typical scattering angle can be computed analogous to equation \ref{eqn:theta_sc}, by noting that the stochastic phase fluctuates by $\sim 1$\,rad over a transverse extent equal to the diffractive scale. This gives $\partial \phi/\partial r\sim 1/r_{\rm diff}$, and the characteristic scattering angle becomes
\begin{equation}
\theta_{\rm sc}=\frac{\lambda}{2\uppi r_{\rm diff}},
\end{equation}
or
\begin{equation}
\theta_{\rm sc} \sim 63\,\upmu{\rm as}\, \left(r_{\rm c}/1\,{\rm pc} \right)^{-1}\lambda_{30}^{11/5} N_{e,17}^{6/5}\left(f_{\rm a}/10\right)^{3/5}
\label{eqn:sc_ensemble}
\end{equation}

The characteristic temporal broadening (equation \ref{eqn:tau}) is much larger than that seen in the Galactic WIM:
\begin{eqnarray}
\Delta\tau &\sim& 0.4\,{\rm ms}\, \left(r_{\rm c}/1\,{\rm pc}\right)^{-2}\lambda_{30}^{22/5}N_{e,17}^{12/5}\left(f_{\rm a}/10\right)^{6/5}\nonumber \\ &&\times\left( \frac{D_{ls}D_l/D_s}{1\,{\rm Gpc}}\right).
\label{eqn:tau_ensemble}
\end{eqnarray}

\subsection{Dependence on halo mass and impact parameter}
\begin{figure*}
\centering
\includegraphics[width=0.7\linewidth]{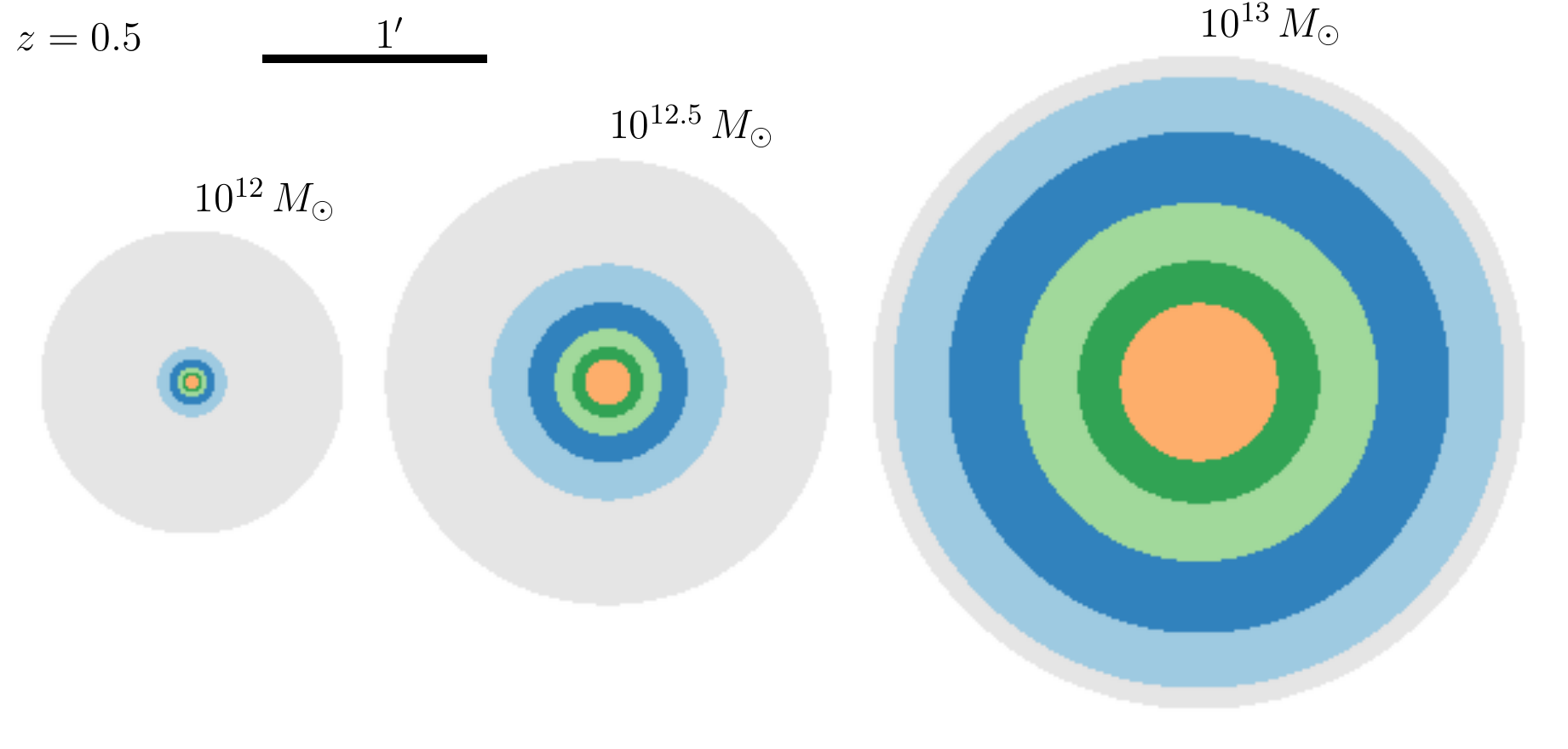}\\
\includegraphics[width=0.7\linewidth]{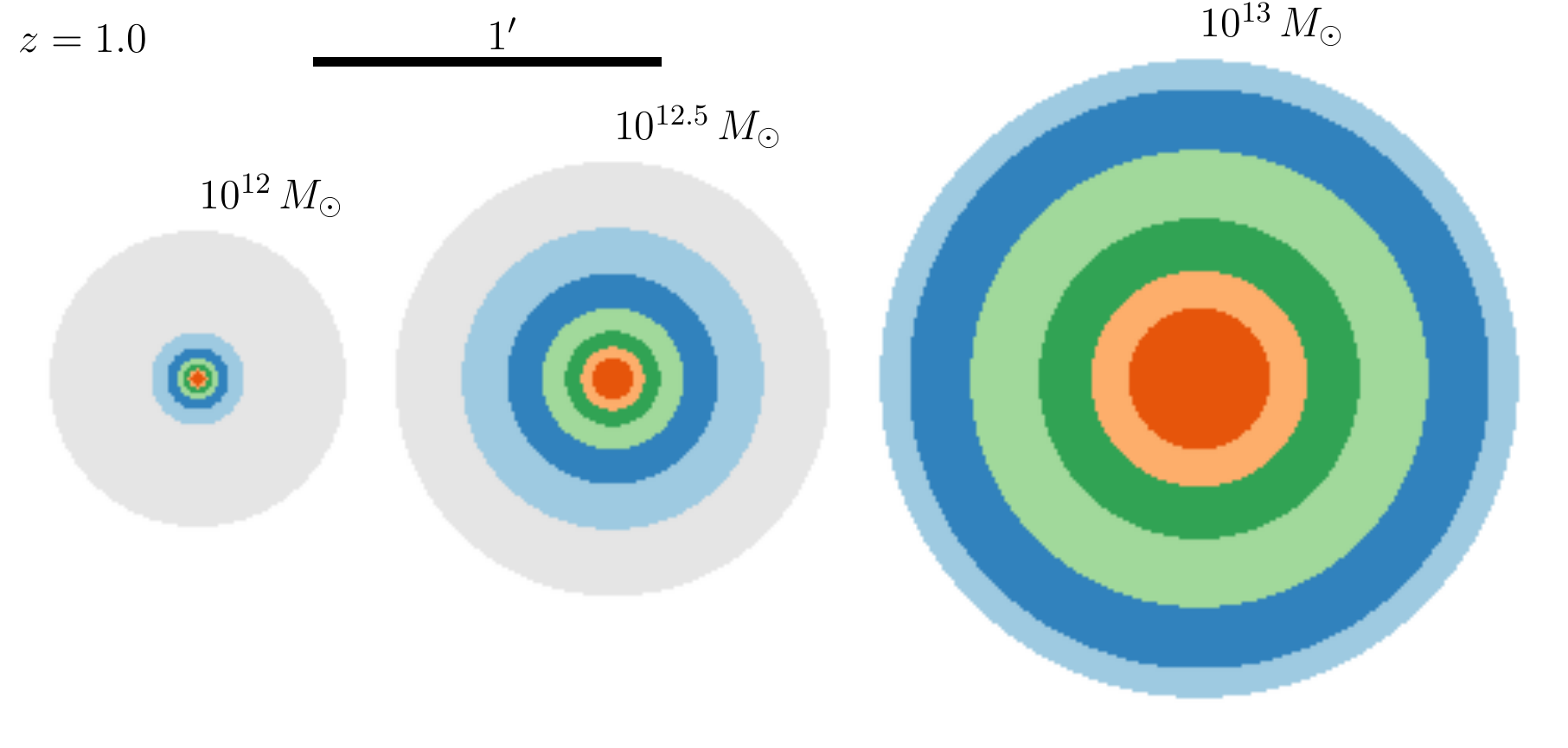}\\
\includegraphics[width=0.7\linewidth]{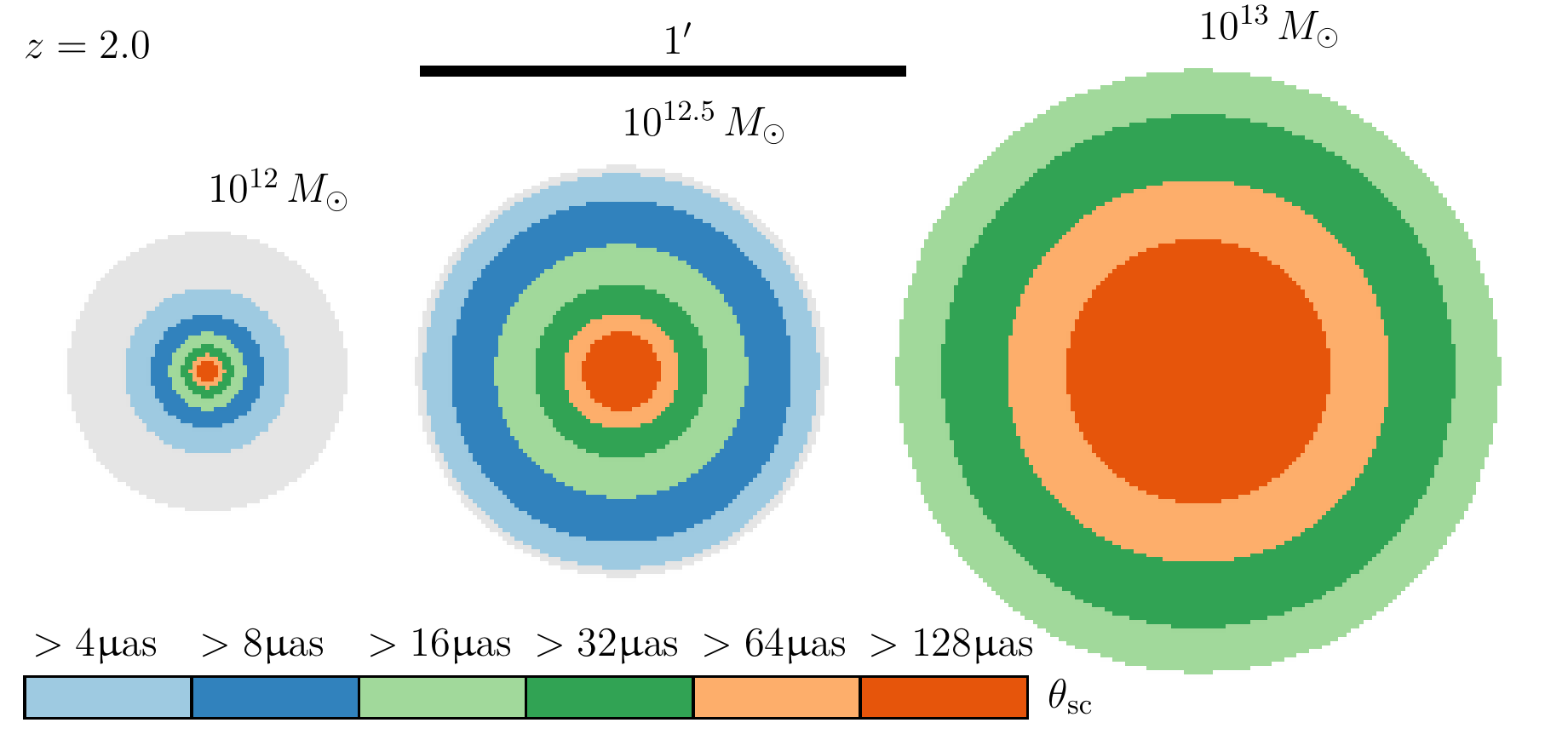}
\caption{To-scale cartoons showing the relative amount of projected sky area within which the scattering angle exceeds the value given by the colour code, for haloes of varying mass ($10^{12}$, $10^{12.5}$ and $10^{13}\,M_{\odot}$). Each row corresponds to a different halo redshift. Scattering strength is parametrized as the characteristics ray deflection angle, $\theta_{\rm sc}$ at an observed wavelength of $\lambda = 30$\,cm (observed size of the scattering disc is $\theta_{\rm ap} = \theta_{\rm sc}\,D_{\rm ls}/D_{\rm s}$). The gray background marks the virial extent of the halo.  \label{fig:scat_cartoon}}
\end{figure*}

\begin{figure*}
\includegraphics[width=\linewidth]{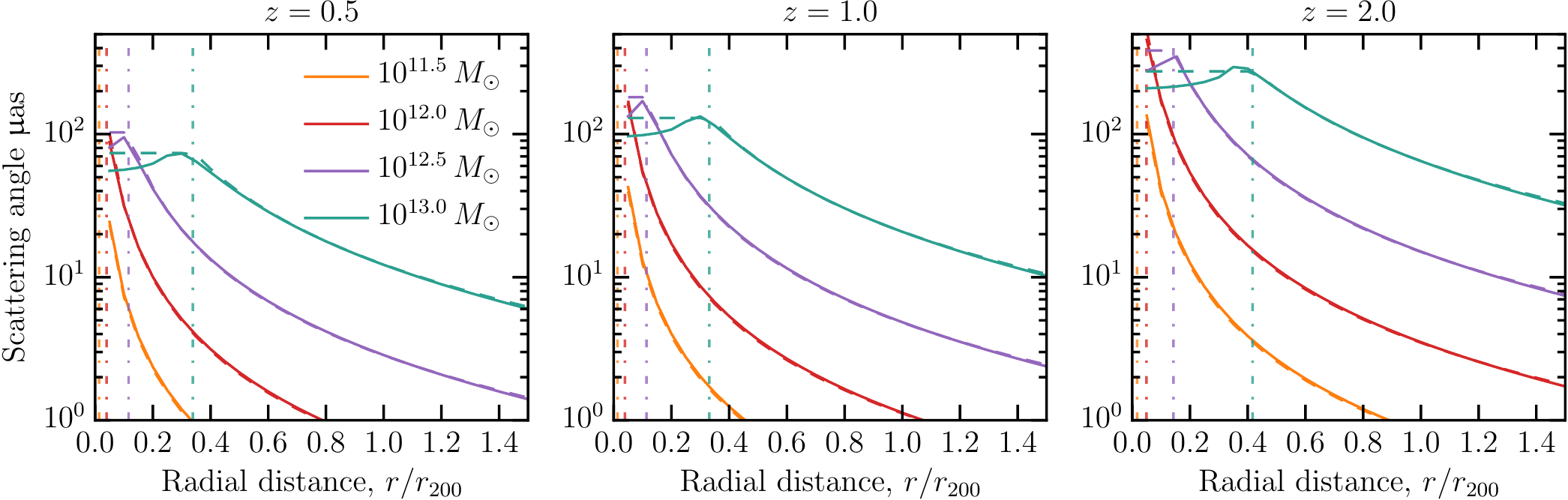}
\caption{Solids lines: plots of scattering angle, $\theta_{\rm sc}$ versus impact parameter, $b_{200}$ (same data as in Fig. \ref{fig:scat_cartoon}) for varying halo mass. Dashed lines: analytical approximation from equation \ref{eqn:halo_theta}. The vertical dot-dashed lines mark $1.5\,r_{\rm ss}$ (defined in equation \ref{eqn:zeta_approx}) where the scattering is expected to saturate due to self-shielding against the ionizing IGM background radiation.\label{fig:sc_vs_b}}
\end{figure*}
Lets us now evaluate the scattering angle and temporal broadening when the sight-line passes a halo of a given mass at some redshift with an impact parameter $b$. Such a ray will pass through cloudlets at varying radii which will possess varying scattering strengths. Ideally one would evaluate the integral $\int {\rm d}y \partial D_\phi(\Delta r,y)/\partial y$ along the ray-path within the halo with $y$ as the affine parameter. However, scattering is dominated by the densest\footnote{By densest, we imply largest $f_{\rm v}$ and smallest $r_{\rm c}$.} part of the halo along the ray path. We can therefore obtain reasonably accurate values for the scattering parameters by assuming that a ray with impact parameter $b$ traverses a distance of $b$ along a cloudlet ensemble with volume filling factor $f_{\rm v}(b)$ and cloudlet radius of $r_{\rm c}(b)$. The areal covering factor along such a ray is $f_{\rm a}(b) = f_{\rm v}(b)\,b/r_{\rm c}$. Following a proceeder similar to that in \S\ref{subsec:ensemble}, we obtain an expression for the diffractive scale at impact parameter $b$:

\begin{equation}
r_{\rm diff}(b_{200}) = \left(2\lambda^2N^2_er^2_e\right)^{-3/5}\,\left[r_{\rm c}(b)\right]^{8/5} \,\left[f_{\rm v}(b) b\right]^{-3/5}.
\end{equation}
Substituting $r_{\rm c}(b) = N_eT/[2P(b)]$ and employing the halo properties from \S\ref{sec:halo_properties}, we get

\begin{eqnarray}
\label{eqn:halo_rdiff}
\frac{r_{\rm diff}(b)}{10^{11}\,{\rm cm}} &=& 3.5\left(\frac{\lambda_{30}}{1+z} \right)^{-1.2}\left(\frac{f_{\rm v}}{10^{-4}} \right)^{-0.6}\left(\frac{f_{\rm CGM}}{0.3}\right)^{-1.6}\nonumber \\ &&\times M_{12}^{-1.27} h^{-3.87}(z)b_{200}^{1.68}\nonumber \\ &&\textrm{for $r_{\rm shock}>b_{200}>1.5r_{\rm ss}$,} \\
&=& 0.17\lambda_{30}^{-1.2}\left(\frac{f_{\rm v}}{10^{-4}}\right)^{0.34}\left(\frac{f_{\rm CGM}}{0.3}\right)^{0.27}M_{12}^{0.29}\nonumber\\
&&\times h^{0.48}(z)\,(1+z)^{-3.47}\,\,\textrm{for $b_{200}<1.5r_{\rm ss}$,}
\end{eqnarray}
where we have enforced the saturation of $r_{\rm diff}$ due to self-shielding. The saturation radius of $1.5r_{\rm ss}$, instead of simply $r_{\rm ss}$ (see equation \ref{eqn:zeta_approx}) was chosen to match $r_{\rm diff}$ versus $b$ profiles obtained from full numerical integration of $\partial D_\phi(y)/\partial y$ along the ray path in the halo. 

The corresponding scattering angle is
\begin{eqnarray}
\label{eqn:halo_theta}
\frac{\theta_{\rm sc}(b)}{\upmu{\rm as}} &=& 2.5\,\left(\frac{\lambda_{30}}{1+z} \right)^{2.2}\left(\frac{f_{\rm v}}{10^{-4}} \right)^{0.6}\left(\frac{f_{\rm CGM}}{0.3}\right)^{1.6}M_{12}^{1.27}\nonumber \\
&&\times h^{3.87}(z)b_{200}^{-1.68}\,\, \textrm{for $r_{\rm shock}>b_{200}>1.5r_{\rm ss}$,}\nonumber \\
&=& 50\,\,\lambda_{30}^{2.2}\left(\frac{f_{\rm v}}{10^{-4}}\right)^{-0.34}\left(\frac{f_{\rm CGM}}{0.3}\right)^{-0.27}M_{12}^{-0.29}\nonumber\\
&&\times h^{-0.48}(z)\,(1+z)^{2.47}\,\,\textrm{for $b_{200}<1.5r_{\rm ss}$.}
\end{eqnarray}
The apprent size of the scattering disc is $\theta_{\rm ap} = \theta_{\rm sc}D_{ls}/D_s$. Finally, the temporal broadening timescale is
\begin{eqnarray}
\label{eqn:halo_tau}
\frac{\Delta\tau}{\rm ms} & = & 7.6\times 10^{-3} \left(\frac{\lambda_{30}}{1+z} \right)^{4.4}\left(\frac{f_{\rm v}}{10^{-4}} \right)^{1.2}\left(\frac{f_{\rm CGM}}{0.3}\right)^{3.2}\nonumber \\
&&\times M_{12}^{2.54} h^{7.74}(z) b_{200}^{-3.36}\left(\frac{D_{\rm eff}}{1\,{\rm Gpc}}\right)\nonumber \\&&\textrm{for $r_{\rm shock}>b_{200}>1.5r_{\rm ss}$,}\nonumber \\
&=& 3\, \lambda_{30}^{4.4}\left(\frac{f_{\rm v}}{10^{-4}}\right)^{-0.68}\left(\frac{f_{\rm CGM}}{0.3}\right)^{-0.54}M_{12}^{-0.58}\nonumber \\
&&\times h^{-0.96}(z)\,(1+z)^{4.94} \left(\frac{D_{\rm eff}}{1\,{\rm Gpc}}\right)\nonumber \\ &&\textrm{for $b_{200}<1.5r_{\rm ss}$.}
\end{eqnarray}
where the effective distance is $D_{\rm eff} = D_{ls}D_l/D_s$. With the above equations, we can now compute the optical depth to scattering for any halo mass function. Figure \ref{fig:scat_cartoon} shows a to-scale depiction of the scattering properties and projected sizes of haloes of various masses and redshifts. Figure \ref{fig:sc_vs_b} compares the analytical approximation to the scattering angle with the result of (a) numerically solving the equilibrium neutral fraction at each location in the halo and then (b) numerically integrating the phase structure function along the CGM sight-line. The agreement is good and we will use equation \ref{eqn:halo_rdiff} to \ref{eqn:halo_tau} to compute the statistics of scattering by a cosmic distribution of haloes in \S4.

\subsection{The impact of granularity}
Before we extent the formalism to account for scattering from multiple halos, we pause to appreciate the impact of pc-scale cloudlet structure on CGM scattering. Consider a $10^{12}\,M_\odot$ halo at $z=1$ with a cool-gas volume fraction of $f_{\rm v}=10^{-4}$, and a fraction $f_{\rm CGM}=0.3$ of baryons in the CGM. The column density of the hot-phase gas would be $N_e\approx 2\times 10^{19}$. If this gas were fully turbulent with an outer scale of $r_{200}\approx 140\,{\rm kpc}$, then its diffractive scale at $\lambda = 30\,{\rm cm}$ is $r_{\rm diff}\approx 4\times 10^{13}\,{\rm cm}$. The diffractive scale due to scattering by cloudlet in our formalism is $r_{\rm diff} \approx 2\times 10^{11}\,{\rm cm}$--- about two orders of magnitude smaller. Hence even though the cool gas only has a volume fraction of $10^{-4}$, it scatters radio waves though a characteristic angle that is two orders of magnitude larger. This is a direct result of the small-scale granularity of cool gas in the cloudlet model considered here. In other words, radio-wave scattering is highly sensitive to the small-scale fluctuations in gas density.

\section{Discussion and summary}
We will now discuss the observable impact of scattering in the CGM. To do so, we first predict the scattering properties of an ensemble of haloes. 
\subsection{Scattering in a $\Lambda$CDM Universe}
\begin{figure}
\centering
\includegraphics[width=\linewidth]{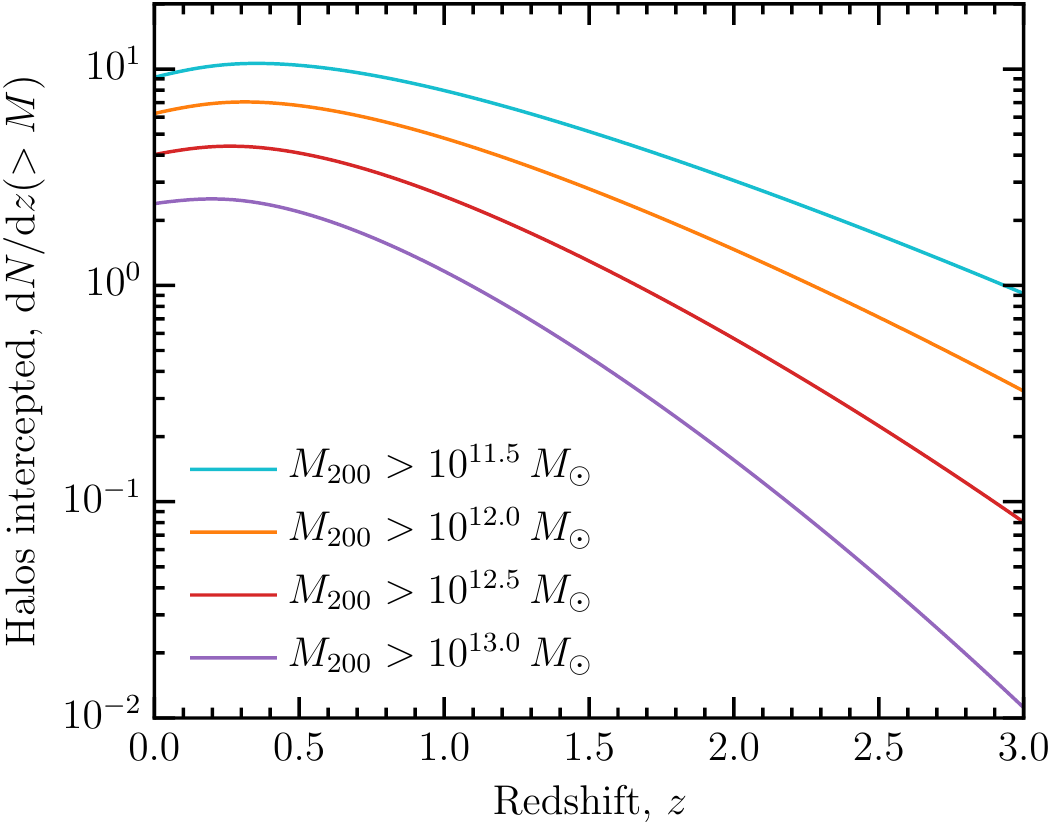}
\caption{Number of virial intercepts by haloes with mass in excess of value shown in legend per unit redshift bin. Halo mass functions were computed using the program of \citet{hmf}, and the footprint of each halo extended to a radius of $r_{\rm shock}=1.5\,r_{200}$ (see eqn. \ref{eqn:r200}).\label{fig:omega}}
\end{figure}

We assume that the volume fraction of cool gas is redshift independent. The halo scattering properties however remain redshift dependent due to the evolution of virial pressure and halo number counts with redshift. We use the halo mass function calculator of \citet{hmf} to compute ${\rm d}N(z,M)/{\rm d}M$--- the co-moving volume density of haloes of with mass in an infinitesimal interval ${\rm d}M$ about $M$, at redshift $z$. Fig. \ref{fig:omega} shows the ensuing numbers of haloes larger than a mass shown in the legend, that are intercepted (within their virial shock) by an average sightline through the Universe. We find that nearly all sight-lines out to $z\sim 1$ pass within the virial radius of a $10^{13}\,M_\odot$ halo, and $\sim~$ten $10^{11}\,M_\odot$ halo. Because larger haloes (a) condense out of the Hubble flow at later times and (b) possess smaller virial radii at higher redshift, the number of intercepts for any given mass range rise up to $z\sim 1$ and decline thereafter.

\begin{figure*}
\centering
\begin{tabular}{ll}
\includegraphics[width=0.5\linewidth]{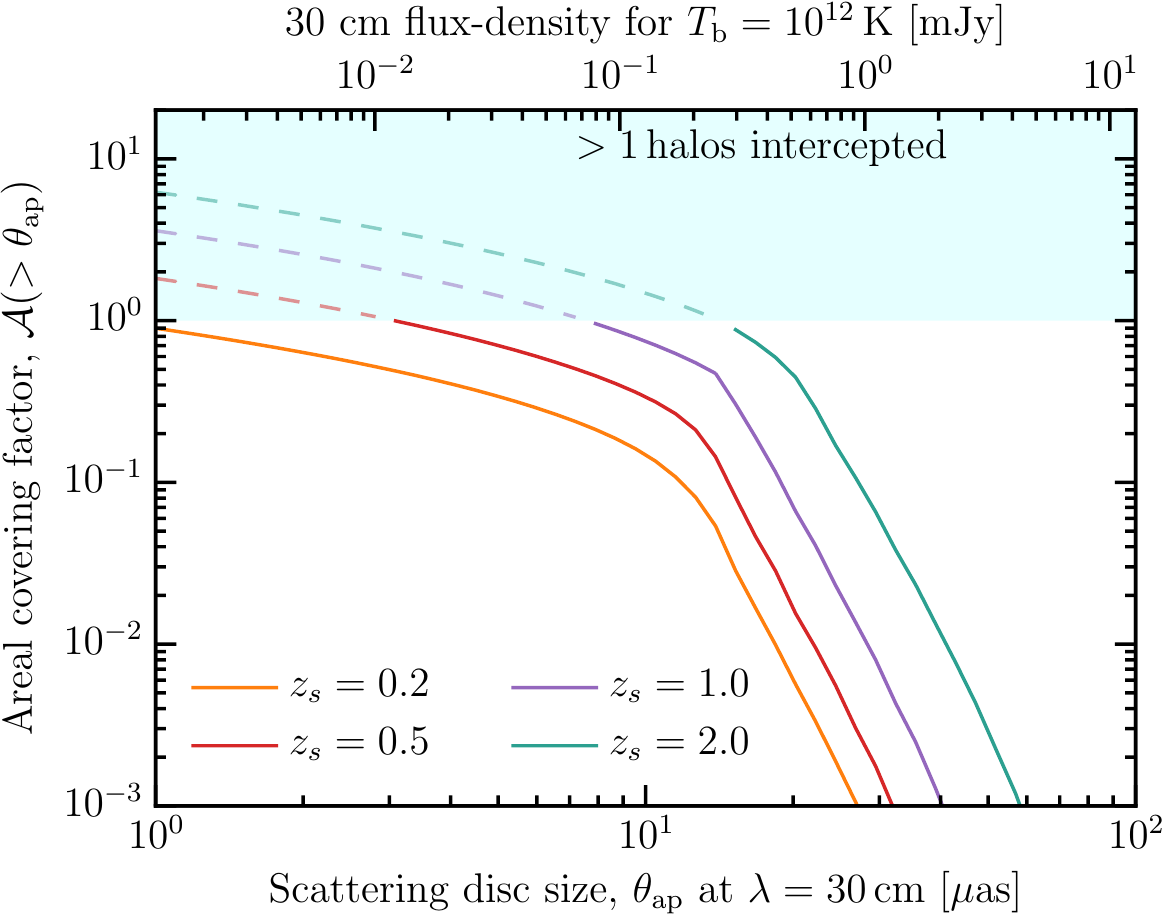} & 
\includegraphics[width=0.48\linewidth]{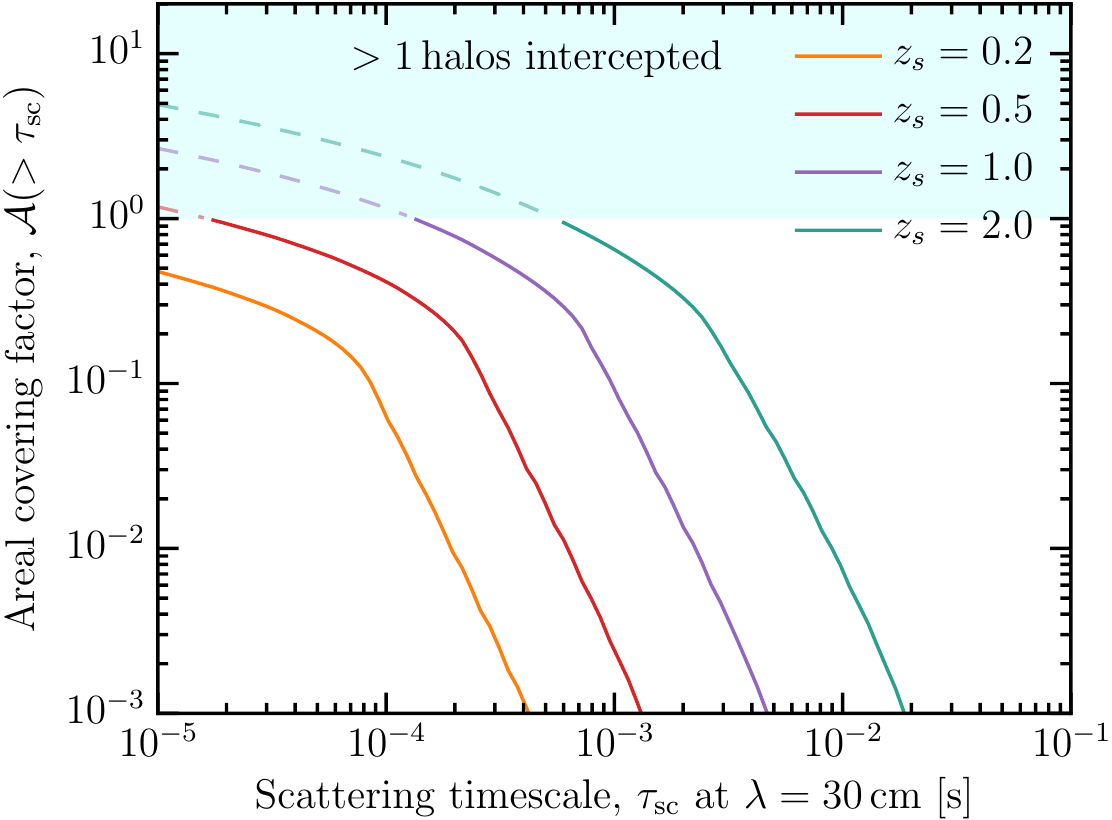} \\
\end{tabular}
\caption{Left: Areal covering factor of sight-lines with scattering angle in excess of value on x-axis. Right: Same as left, but for scattering timescale. A covering factor $>1$ implies there is more than one intervening halo whose scattering individually exceeds the x-axis value. Assumed parameter values are $f_{\rm v}=10^{-4}$, $f_{\rm CGM}=0.3$, $\Gamma =10^{-13}\left[(1+z_l)/1.2\right]^5\,$s$^{-1}$\label{fig:theta_tau_cdm}}
\end{figure*}

Consider a radio source at redshift $z_s$. The statistics of the scattering timescales from all intervening haloes at redshift $z_l<z_s$ can be computed as follows. We pick a scattering timescale $\tau_{\rm sc}$, and for each halo mass and redshift bin, we compute the impact parameter $b_{\rm max}(M,z_l,z_s,\tau_{\rm sc})$ below which the scattering timescale exceeds $\tau_{\rm sc}$ via equation \ref{eqn:halo_tau}. The projected area is $\uppi b_{\rm max}^2/D_l^2$. There are ${\rm d}N(M,z_l)/{\rm d}M$ haloes per unit volume that contribute to the scattering. The areal covering factor of sightlines whose scattering timescale exceed $\tau_{\rm sc}$ is therefore
\begin{eqnarray}
\mathcal{A}(>\tau_{\rm sc}) &=& \int {\rm d}^2V(z)\,\int_{M_{\rm min}}^{M_{\rm max}}\,{\rm d}M\, \frac{{\rm d}N(M,z_l)}{{\rm d}M}  \nonumber \\ && \times \frac{\uppi b_{\rm max}^2(M,z_l,z_s,\tau_{\rm sc})}{D_l^2}
\end{eqnarray}
where $M_{\rm min}$ and $M_{\rm max}$ are the mass range of interest, assumed to be $10^{11.5}\,M_\odot$ and $10^{13.5}\,M_\odot$ respectively, and ${\rm d}^2V(z)$ is the co-moving volume element at redshift $z$ given by:
\begin{equation}
{\rm d}^2V(z) = c^3\,\frac{\left(\int_0^z\,{\rm d}z\prime / H(z\prime)\right)^2 }{H(z)}\,{\rm d}z\,{\rm d}\Omega
\end{equation}

We note that $\mathcal{A}(>\tau_{\rm sc})$ can be larger than unity which indicates that there is more than one halo along the sight-line whose `stand-alone' scattering strength exceeds $\tau_{\rm sc}$.
An identical procedure can be followed for any other scattering parameter such as the apparent size of the scattering disc. 
Figure \ref{fig:theta_tau_cdm} shows the apparent angular size of the scattering disc and the scattering timescale calculated using the above prescription. The figure shows that most sight-lines out to $z_{\rm s} = 1$ suffer angular broadening of at least $\sim 8\,\upmu$as and temporal broadening of at least $\sim 0.1$\,ms. The scattering for $z_{\rm s}\gtrsim 0.2$ happens due to many intervening haloes. To understand their effect, we must compute the average scattering angle. 

\begin{figure*}
\centering
\includegraphics[width=\linewidth]{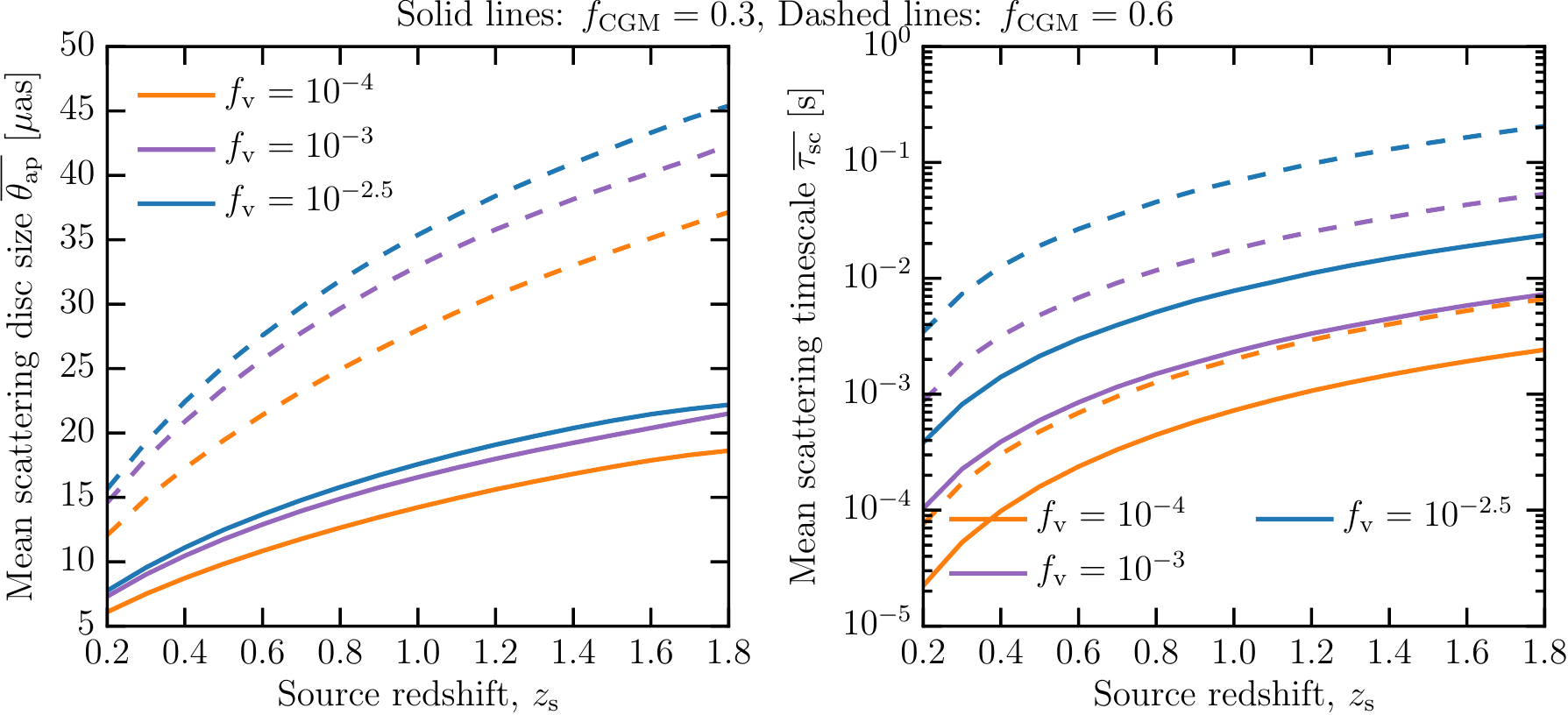}
\caption{Mean angular (left) and temporal (right) broadening as a function of source redshift. Different lines show variation in cool gas volume fraction $f_{\rm v}$ and the fraction of cosmic baryons in the CGM, $f_{\rm CGM}$. A photoionization background of $\Gamma(z) = 10^{-13}\left[(1+z)/1.2\right]^5$\,s$^{-1}$ has been assumed.\label{fig:meantau}}
\end{figure*}

To first order, the scattering angle due to multiple scattering `screens' add in quadrature, and the scattering timescale add linearly \citep[][their Appendix A]{blandford1985}. We can therefore compute the mean scattering timescale as 
\begin{equation}
\overline{\tau_{\rm sc}} = -\int_0^{\infty}\,{\rm d}\tau\,\tau\,\frac{{\rm d} \mathcal{A}(>\tau)}{{\rm d} \tau}
\end{equation}
where (the negative of) the differential in the integrand returns the probability density function of $\tau$ and the integral therefore evaluates to the expected value of $\tau$. The mean size of the scattering disc is similarly
\begin{equation}
\overline{\theta_{\rm ap}} = \sqrt{-\int_0^{\infty}\,{\rm d}\theta \,\theta^2\,\frac{{\rm d} \mathcal{A}(>\theta)}{{\rm d} \theta}}
\end{equation}
Figure \ref{fig:meantau} shows the mean temporal and angular broadening this computed as a function of source redshift for different values of cool-gas volume fraction $f_{\rm v}$ and fraction of baryons in the CGM, $f_{\rm CGM}$. The fractional sample variance on the mean is driven in large part by the Poisson fluctuations in the number of intercepted haloes. Based on Fig. \ref{fig:omega}, the fractional variation is of order unity for $z_{\rm s}\lesssim 0.2$ and reduces to few tens of per-cent by $z_{\rm s}\sim 1$.   order unity for $z_{\rm s}\lesssim 0.2$ and decreases to tens of per-cent by $z_{\rm s}\sim 1$.
\subsection{How can CGM scattering be observed?}
Figure \ref{fig:meantau} shows that sources at $z_s\gtrsim 1$ are scatter broadened to typical angular size of $\sim 20\,\upmu$as and in timescale to about $\gtrsim 1$\,ms, at a wavelength of $\lambda = 30\,$cm. Despite the considerable uncertainty in parameters affecting CGM scattering (specifically $f_{\rm v}$ and $f_{\rm CGM}$), let us take these numbers as a fiducial test case, to understand the observational manifestation of CGM scattering.

\subsubsection{Refractive and diffractive scales}
\begin{table}
\caption{A comparison of characteristic angular and time-scales for scattering in the CGM (this work) and the Galactic warm ionized medium \citep{walker1998}. A refractive scale of $\theta_{\rm ap}=20\,\mu$as at $\lambda = 30$\,cm has been assumed for the CGM contribution (see Fig. \ref{fig:theta_tau_cdm}). \label{tab:scale}}
\begin{tabular}{lll}
\hline \\
\textbf{Parameter} & \textbf{CGM} & \textbf{MW}\\ && \textbf{(high lat.)} \\ \hline \\
Transition wavelength ($\lambda_{\rm tran}$)	     & $0.3\,$cm	        & 3.75\,cm\\
Length-scale	                                     & $1\,$Gpc	      	    & $1\,$kpc\\
Fresnel scale at $\lambda_{\rm tran}$	             & $10^{-3}\,\upmu$as     & $3\,\upmu$as\\
Diffractive angular scale ($\lambda = 30\,$cm)       & $10^{-6}\,\upmu$as     & $0.25\,\upmu$as\\
Refractive angular scale ($\lambda = 30\,$cm)        & $20\,\upmu$as          & $0.3\,$mas\\
Diffractive time scale ($\lambda = 30\,$cm)          & $5$\,min/$0.5$\,s$^{\dagger}$      & $10$\,min \\
Refractive timescale$^{\dagger}$ ($\lambda = 30\,$cm)            & $180$\,yr/$0.3$\,yr$^\dagger$             & $8\,$d\\
Temporal pulse broadening ($\lambda = 30\,$cm)		 & $1\,$ms				& $0.17\,\upmu$s\\ \hline
\end{tabular}
$^\dagger$ For the CGM case, the refractive timescales are quoted for two cases: (a) when the transverse velocity is $500\,$km\,s$^{-1}$ and (b) when the transverse velocity equals the speed of light (relevant for relativistically moving sources).\\
\end{table}

\begin{figure}
\centering
\includegraphics[width=\linewidth]{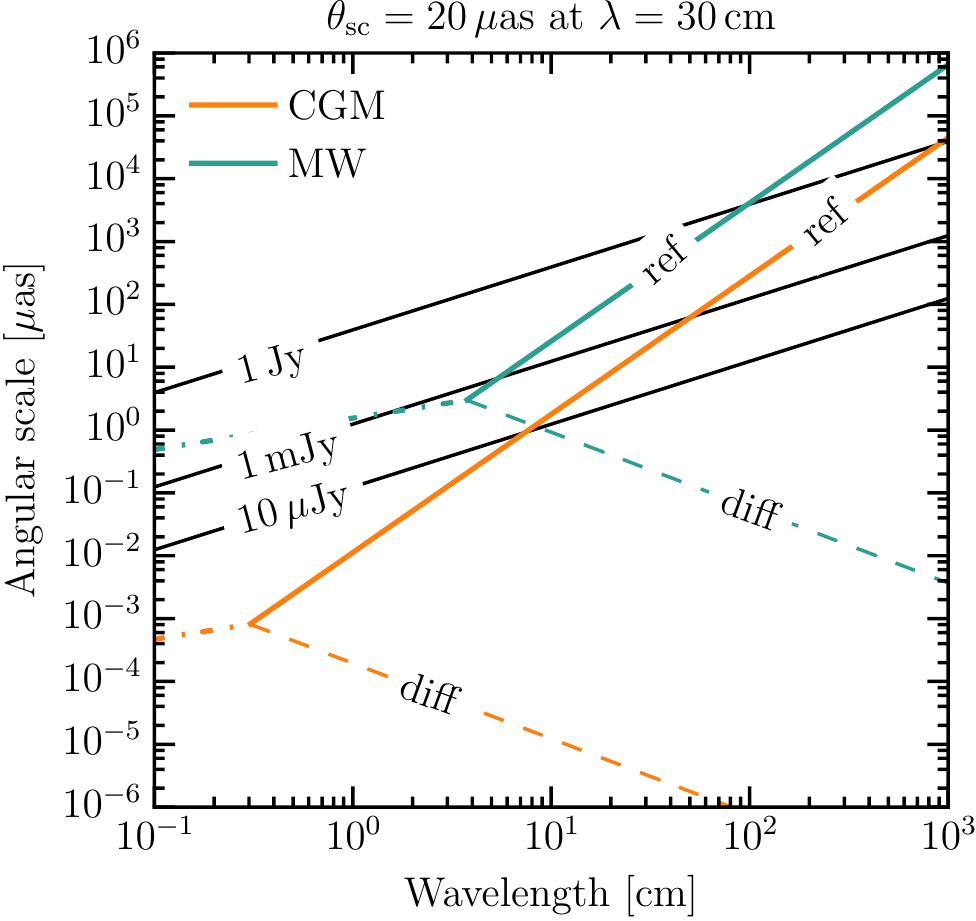}
\caption{Plot showing the typical refractive and diffractive scales in the CGM (green lines) and Milky-way (orange lines) for a source at $z\gtrsim 1$. In either set of curves, the dot$-$dashed line ($\propto \lambda^{1/2}$) shows the weak scattering regime below the transition wavelength. In the strong scattering regime, the solid ($\propto \lambda^{11/5}$) and dashed ($\propto \lambda^{-6/5}$) curves show the refractive and diffractive scales respectively. The solid black lines ($\propto \lambda$) show the intrinsic angular size of an (incoherent) synchrotron source with a brightness temperature of $10^{12}\,$K, and flux-density values given in the in-line labels. A length-scale of $1$\,Gpc has been assumed to convert all physical scales to angular scales. \label{fig:scale}}
\end{figure}

We first summarize the relevant aspects of two regimes of scattering: diffractive and refractive.\footnote{Also called fast and slow scintillation respectively. See \citet{rickett1984,goodman1985} for further details.} Diffractive effects manifest on scales given by $\theta_{\rm diff} = r_{\rm diff}/D_{\rm l}$ on which individual speckles form. The ensemble of speckles form a scattering disc over the refractive scale given by $\theta_{\rm ap} = \theta_{\rm sc}D_{\rm ls}/D_{\rm s} = \lambda /(2\uppi \theta_{\rm diff})\,D_{\rm ls}/(D_{\rm s}D_{\rm l})$. Because $r_{\rm diff}$ evolves as $\lambda^{-6/5}$, the diffractive and refractive scales evolve as $\theta_{\rm diff}\propto \lambda^{-6/5}$ and $\theta_{\rm ap} \propto \lambda^{11/5}$. 
When the scattering is too weak to form speckles, the apparent size of a point-like source is set by the size of the first Fresnel zone given by $\theta^2_{\rm f} = \lambda/(2\pi)\,D_{\rm ls}/(D_{\rm l}D_{\rm s})$. It is trivial to show that $\theta^2_{\rm f} = \theta_{\rm diff}\theta_{\rm ap}$ and that all three angular scales are equal to one another at the transition wavelength: $\lambda_{\rm tran} = 2\uppi  r^2_{\rm diff}\,D_{\rm s}/(D_{\rm l}D_{\rm ls})$. Below the transition wavelength, scattering is weak and manifests as weak flux-density modulation due to plasma density fluctuations that focus and de-focus the electromagnetic wavefront on the Fresnel scale. Above the transition wavelength, diffractive flux-density modulations results from fluctuations in the position and brightness of speckles that interfere at the observer, while refractive modulations result from focussing and de-focussing of the entire speckle ensemble. The above discussion applies to point-like sources. Refractive and diffractive scintillation of extended sources are rapidly `washed out' as the intrinsic source size exceeds the diffractive and refractive scales respectively. Figure \ref{fig:scale} and table \ref{tab:scale} summarize the angular and time-scale of scintillation in the Galactic WIM at high Galactic latitudes and the corresponding CGM values for our fiducial test case.

\subsubsection{Incoherent synchrotron sources}
Let us first consider incoherent optically-thin synchrotron sources with a characteristic brightness temperature of $10^{12}\,$K. Although the flux-density of such sources is strongly modulated by refractive effects for $\lambda\gtrsim 0.3\,$cm, the time-scale over which these modulations manifest in light-curves is too large to be of practical interest. More importantly, even sources as faint as $10\,\upmu$Jy are too large for refractive modulations to be observable at $\lambda\lesssim 10\,$cm, whereas at $\lambda\gtrsim 10\,$cm, the flux-density modulations are dominated by the Galactic WIM. Hence the influence of CGM scattering will be difficult to identify observationally using incoherent synchrotron sources. This conclusion also serves as an essential `sanity-check' in that, our postulated existence of significant CGM scattering does not violate the large existing body of work on scintillation of incoherent synchrotron sources (active galactic nuclei and gamma-ray burst afterglow for e.g.) that only consider flux-modulations from Galactic scintillation.

There is however a narrow parameter range where CGM scattering may be discerned from Galactic scattering in weak $<10\,\upmu$Jy-level sources. Consider the $3\,{\rm cm}\lesssim \lambda \lesssim 10\,{\rm cm}$ regime in Fig. \ref{fig:scale}. In the absence of CGM scattering, weak sources may be small enough to display diffractive scintillation in the Galaxy which could be observed as modulations in the radio spectrum of sources on scales of $\Delta\nu/\nu \approx \left(\nu/\nu_0\right)^{17/5}$ \citep[][their \S 3.2.2]{walker1998}. However, these scintillations will be quenched in the presence of angular broadening of the source in intervening CGM which could push the apparent source size above the Galactic diffractive scale. Given the large uncertainty in predictions for CGM scintillation parameters it is difficult to accurately predict where this wavelength window exists for a given sight-line. A targeted survey of sources along sight-lines at varying impact parameters (which would vary the transition frequency in Fig. \ref{fig:scale}) may be a fruitful avenue to explore. Assuming a characteristic coherence scale of $\Delta\nu\sim 1\,$GHz, $\tau\sim 1$\,hr for Galactic diffractive scintillation, a system temperature of $30$\,K, aperture efficiency of $60\,$per-cent, such an experiment would require a collecting area well in excess of $\sim 10^5$\,m$^{2}$ which is barely within reach of existing radio telescopes.

We have also considered early radio emission from gamma-ray bursts (GRBs), which can have higher brightness temperatures at early times than blazars, owing to their ultrarelativistic velocities. They can therefore be brighter and easier to measure while still at small angular sizes, and are consequently observed to show interstellar scintillation in their first days at $\sim 5\mbox{GHz}$ \citep{granot2014}. Before deceleration to Lorentz factor $\Gamma<1/\theta_j$ (before the ``jet break'' for a jet of opening half-angle $\theta_j$), the projected source angular size $\theta$ at (earth) time $T$ after explosion of a GRB at redshift $z$ is $\theta\sim 2cT\Gamma/D_M(z)$, where $D_M(z)=D_A(z)(1+z)$ is the proper motion distance, and $D_A(z)$ the angular diameter distance. The Blandford-McKee blast wave of the ultrarelativistic shock moving into a medium of uniform external density $\rho_0$ has radius $R\simeq 2cT\Gamma^2/(1+z)$ and explosion energy per unit solid angle $E/\Omega\simeq \rho_0 R^3 \Gamma^2 c^2$, which gives $\Gamma\simeq 9(E_{iso,53}/n_0)^{1/8}(T/[(1+z)\mbox{day}])^{-3/8}$, where $E=10^{53}\mbox{erg}(\Omega/4\pi)E_{iso,53}$ and $n_0$ is the external density in $\mbox{cm}^{-3}$ \citep[][cf.]{granot2002}. At $D_M(z=1)=3.3\mbox{Gpc}$, $\theta= [0.2,\,1,\,4]\,\upmu\mbox{as}$ at $T=[0.1,\,1,\,10]\,\mbox{d}$.  Thus at $\lambda<4\mbox{cm}$ (the transition wavelength below which Milky Way scintillation becomes unimportant), the GRB will be smaller than our fiducial scattering angle $\theta=20\upmu\mbox{as}(\lambda/30\mbox{cm})^{11/5}<0.25\upmu\mbox{as}$ for less than 0.1~day. During this time, the scintillation timescale will be set by the rapidly expanding source, expanding across the refractive screen at a projected speed of $\sim \Gamma c D_l/D_s$. This is many times $c$ for our cosmological lenses with $D_l\sim 0.5D_s$ (but less than $1\,\mbox{km s}^{-1}$ for Milky Way interstellar plasma at $D_l\sim 100\mbox{pc}$, so Milky Way scintillation timescales are dominated by gas motions in the Milky Way, not the apparent source expansion).  The refractive scintillation timescale is thus the same as the timescale for the source to expand to a size larger than the refractive scale ---i.e., the source will have only about 1 speckle before becoming too large to display refractive scintillation.  This would be difficult to convincingly detect in a GRB.  

We thus turn to the most promising class of sources.

\subsubsection{Coherent sources}
Fast radio bursts \citep[FRB; ][]{lorimer2007} are the only known class of  coherent emitters at cosmic distances of interest to CGM scintillation. Extragalactic mega-masers are known to scintillate due to the Galactic turbulence \citep{argon1994}. However, even if they are compact enough to show diffractive scintillation in intervening CGM, the interpretation is clouded by the possibility of intrinsic variability \citep[see e.g. ][]{greenhill1997}, and we will not consider them here. FRBs are $\lesssim 1$\,ms duration bright ($\sim 1\,$Jy) radio bursts of extragalactic origin. At least one FRB is known to repeat \citep{spitler2016} which is the only FRB to have been securely localized, and resides in a galaxy with redshift $z_s=0.193$ \citep{chatterjee2017,tendulkar2017}. However, if most of the observed plasma dispersion is apportioned to the intergalactic medium, then the known populations of FRBs with dispersion measures ${\rm DM}\sim 500-2000\,$pc\,cm$^{-3}$ \citep{petroff2016,ravi2017} originate at redshifts of $z\sim 0.5-2$. In this redshift range, the spectra of quasars show absorption systems, e.g. in Mg II, CIV, Lyman limit systems produced in
the halos of one or more intervening galaxies\citep{steidel1988,steidel1994,mathes2017}. Thus signals from cosmological FRBs must also be passing through the cool ionized clumps in the CGM of galaxies.

Based on their $\sim \,$ms duration, FRBs should projected an angular size of $\sim 10^{-6}\,\upmu$as at $D_{\rm s}\sim 1$\,Gpc, even if the emission region travels with apparent superluminal speed with relativistic $\gamma \sim 10^3$. Hence FRBs must display the effects of diffractive (and refractive) scintillation in both intervening CGM and the Galactic WIM. The characteristic  pulse broadening time-scale in the CGM of $\gtrsim 1\,$ms should also be easily distinguishable from the $\lesssim 0.1\,\upmu$s of broadening expected in the Galactic WIM at high latitudes, and a presumably similar amount from the FRB host galaxy. Some FRBs may also originate in dense star-forming regions which may contribute significantly to temporal broadening. CGM scattering can however be distinguished in a population of localized FRBs in two ways. (a) One can attempt a statistical detection of an FRB temporal broadening v/s redshift relationship and constrain the amount of cool gas in the CGM fog (via Fig. \ref{fig:meantau}. (b) The variation of temporal broadening with halo mass and impact parameter can be measured (with significant investment of time on optical spectrographs, comparable to that
invested in quasar studies, e.g.\citep{steidel1994}) and CGM scattering constrained via equation \ref{eqn:halo_tau} and \ref{fig:scat_cartoon}. These appear to be the most promising avenues to directly constrain the fine sub-parsec scale properties of cool gas in the CGM.

With the current absence of a sample of well localized FRBs, we can only make a heuristic comparison between our predictions and data. If a large fraction of the observed FRBs at $\lambda=30\,{\rm cm}$ are indeed at $z\sim 1$ as they dispersion measures suggest \citep{petroff2016}, then based on Fig. \ref{fig:theta_tau_cdm}, the most extreme models with $f_{\rm V}\gtrsim 10^{-3}$ and $f_{\rm CGM}\gtrsim 0.6$ are disfavored. The more moderate models such as ($f_{\rm CGM}=0.3,\,f_{\rm V}=10^{-4}$) or ($f_{\rm CGM}=0.6,\,f_{\rm V}=10^{-4}$) are broadly consistent with the $\sim $ms scale scattering seen in some FRBs if they are at $z\sim 1$. The same models also predict $>1\,{\rm s}$ of scattering at frequencies below $\sim 200\,{\rm MHz}$, making them difficult to detect. This is a plausible explanation for the current non-detection of FRBs at such low frequencies \citep{chawla2017,tingay2015,artemis}.

\subsection{Summary}
In addition to the hot $10^6$\,K halo gas, quasar absorption spectroscopy and fluorescent Ly$\upalpha$ imaging have detected large amounts of cool $10^4\,$K gas in the CGM of $\gtrsim 10^{12}\,M_\odot$ haloes. This was not predicted in canonical galaxy assembly models, but has been accounted for in recent simulations of  cooling instabilities that drive the formation of numerous sub-pc size cloudlets of cool gas. The tiny size of these cloudlets make their spectroscopic or imaging-based detection (and even study via simulations) difficult. We have shown that the pc-scale `granularity' imparted by the small cloudlet size results in a large increase in their radio wave scattering strength. The resulting temporal broadening at $\lambda=30$\,cm of $\sim 10^{-1}-10$\,ms (depending on cool gas volume fraction and fraction of baryons in CGM) far exceeds that expected from the Galactic WIM. This makes their study feasible with fast radio bursts. Identification of our predicted associated temporal broadening in FRBs could revolutionized study of small-scale structure of the
CGM in much the same way that pulsars revolutionized our understanding of sub-au scale structure in the Galactic WIM. We have computed the scattering properties of individual haloes (equations \ref{eqn:halo_rdiff} to \ref{eqn:halo_tau}) as function of halo parameters and redshift, as well as ensemble scattering properties through sight-lines in the Universe (\ref{fig:meantau}). The imprint of CGM scattering on the angular size and scintillation of faint compact radio sources may be difficult to discriminate from scattering in the Galactic WIM. A population of well localized FRBs however, will provide a much more promising avenue to measure the sub-pc scale structure of the CGM. Such a measurement will however have to discriminate between scattering in intervening CGM and other plausible scattering sites such as the circum-burst medium. 

We end by noting that while we have demonstrated the observable scattering effects of cool gas clouds, the precise CGM model considered here is likely simplistic. For instance, \citet{mccourt2016} considered equilibrium cooling rates for collisionally ionized solar-metallicity gas that is optically thin. These and other assumptions \citep[see e.g. \S 4.1 of][]{mccourt2016} likely break-down at least in some parameter ranges of redshift, halo mass and galaxy type. Our {\em method} to compute CGM scattering from small-scale cool-gas clouds presented here can however be readily adapted to future refinement of CGM cool-gas models.  

\section*{Acknowledgements}
HKV is an R.~A. \& G.~B. Millikan fellow of experimental physics and thanks S.~Kulkarni for discussions. ESP's research was funded in part by the Gordon and Betty Moore Foundation through grant GBMF5076. Figs. 1 and 3 were rendered using \texttt{inkscape} and the rest using \texttt{matplotlib}. Numerical computations were carried out in \texttt{python2.7} and employed routines from the \texttt{numpy} package.

\appendix
\onecolumn

\begin{table*}
\centering
\caption{Glossary of  symbols and their implied meaning}
\begin{tabular}{|ll|}
{\bf Symbol}		& {\bf Meaning}\\ \hline \\
$\lambda$		& Observed wavelength\\
$\lambda_{30}$		& $\lambda$ in units of 30\,cm\\
$N_e$			& Electron column density through a single cloudlet\\
$N_{e,17}$		& $N_e$ in units of $10^{17}\,{\rm cm}^{-2}$\\
$r_{\rm c}$		& Radius of a single cloudlet\\
$f_{\rm a}$	& Average number of cloudlets intercepted\\ 
$f_{\rm CGM}$	& Fraction of baryons in CGM (halo mass and redshift independent)\\
$r_{\rm diff}$		& Diffractive scale of plasma inhomogeneities\\
$\theta_{\rm sc}$	& Characteristic wave scattering angle\\
$\Delta\tau$		& Characteristic pulse broadening timescale\\
$r_{200}$		& Radius at which halo density is $200\times$ critical density\\
$M_{200}$		& Mass enclosed within $r_{200}$\\
$M_{12}$		& $M_{200}$ in units of $10^{12}\,M_{\sun}$\\
$b$				& Impact parameter\\
$b_{200}$		& Impact parameter in units of $r_{200}$\\
$b_{\rm c}$		& Radius below which cloudlets self-shield against photoionization\\
$b_{\rm c,200}$	& $b_{\rm c}$ in units of $r_{200}$	\\
$P_{200}$		& Gas pressure at $r_{200}$\\
$T$, $T_4$		& Gas temperature in units of Kelvin and $10^4$\,Kelvin respectively\\
$\alpha$		& Power-law index for variation of gas pressure with radius\\
$\beta$			& Power-law index for variation of cool-gas volume fraction with radius\\
$D_\phi(r)$		& Phase structure function at transverse separation $r$\\
$D_l$			& Observer$-$Lens angular diameter distance\\
$D_s$			& Observer$-$Source angular diameter distance\\
$D_{ls}$		& Lens$-$Source angular diameter distance\\
$r_e$			& Classical electron radius $\approx 2.81794\times 10^{-13}\,$cm.\\ \hline
\end{tabular}
\end{table*}
\section{Phase structure function of a cloudlet ensemble}
Let the two dimensional impact parameters vector be $\bm{b}=[b_x,\,b_y]$, and let us use the notation $|\bm{x}|\equiv x$ for spatial vectors. If there are $f_{\rm a}$ cloudlets intercepted by a ray, then the number of cloudlets intercepted with impact parameters within an interval ${\rm d}b_x\,{\rm d}d_y$ around $[b_x,\,b_y]$ is a Poisson random variable with mean
\begin{equation}
\left<\frac{\partial^2N(b_x,b_y)}{\partial b_x\,\partial b_y}\right> = \frac{f_{\rm a}}{\pi r_{\rm c}^2}. 
\end{equation}
Let $f(\bm{b}) \in [0,\,1)$ be the fractional path-length through a cloudlet. Because $\lambda r_e N_e$ is the maximal phase through a cloudlet, the phase inserted into a ray at impact parameter $\bm{b}$ is therefore $\lambda r_e N_e f(\bm{b})$. With these definitions, the phase structure function at transverse separation $\Delta r$ can be written as the ensemble average:
\begin{equation}
D_\phi(\Delta r) =\left<\left( \int\,{\rm d}b_x\,\int\,{\rm d}b_y\, \frac{\partial^2N(b_x,b_y)}{\partial b_x\,\partial b_y}\, \lambda r_e N_e \left[f(\bm{b})-f(\bm{b+\Delta r}) \right]\right)^2\right>. 
\end{equation}
We now bring in the assumption that cloudlets are randomly distributed. The random variable $\partial^2N(b_x,\,b_y)/\partial b_x\,\partial b_y$ therefore has the properties
\begin{eqnarray}
\left<\frac{{\rm d}^2N(b_x,\,b_y)}{{\rm d}b_x\,{\rm d}b_y} \frac{{\rm d}^2N(b^\prime_x,\,b^\prime_y)}{{\rm d}b^\prime_x\,{\rm d}b^\prime_y} \right> &=& \frac{2f_{\rm a}}{\pi r_{\rm c}^2}\,\,{\rm if}\,b_x=x^\prime_x,\,b_y=b^\prime_y\nonumber\\
&=&0\,\,{\rm otherwise}
\end{eqnarray}
With this, the structure function reduces to
\begin{equation}
D_\phi(\Delta r) = \lambda^2r_e^2N_e^2 \, 2f_{\rm a}\,\times\, \frac{\int\,{\rm d}b_x\,{\rm d}b_y\,\left[ f(\bm{b}) - f(\bm{b}+\Delta\bm{r})\right]^2}{\pi r_{\rm c}^2}.
\end{equation}
The first factor is the variance of the phase accumulated by a ray propagating through the cloudlet ensemble. The second factor, defined as the function $\Psi(.)$ in \S 3, only depends on the internal structure of the cloudlets. For axially symmetric cloudlets, it is only a function of $\Delta r/r_{\rm c}$. It increases from $0$ for $\Delta r =0$ and saturates at $1$ for $\Delta r \geq r_{\rm c}$

\section{Photoionization balance in the galactic fog}
We make the following simplifying assumption in our computation of the photoionization balance: (a) all neutral hydrogen atoms are in the ground state, (b) photons emitted during direct recombinations to the ground state are all reabsorbed `close by' in the halo (so called on-the-spot approximation), (c) only photons close to $\nu_0=3.3\times 10^{15}$\,Hz participate in the ionization balance (d) the free-electrons have a Maxwellian distribution owing to their large elastic scattering cross-section ($\sim 10^{-13}$\,cm$^{-2}$).  

Let $J_{\nu_0}(r)$ be the number of photons at frequency $\nu_0$ per unit area, per unit solid angle, per unit frequency per unit time present in the halo at radial distance $r$ (i.e. $\mathcal{I}_{\nu_0} = J_{\nu_0}/(h\nu_0)$). The boundary condition for photoionization equilibrium is set by the measured/modelled UV photon field in the IGM, which is typically specified as an isotropic photoionization rate:
\begin{equation}
\Gamma({\rm IGM}) = 4\pi\,J_{\nu_0}({\rm IGM})\,a_0 
\end{equation}
where $a_0\approx 6.3\times10^{-18}$\,cm$^{-2}$. \citet{gaikwad2017} determined a photoionization rate of $\Gamma({\rm IGM}) = 10^{-13}\left[(1+z)/1.2\right]^{5}$. This gives the boundary condition of
\begin{equation}
4\pi a_0 J_{\nu_0} = 10^{-13}\left[(1+z)/1.2 \right]^5
\end{equation}

At any location $r$ in the halo, if $\zeta(r)$ is the neutral fraction, then photionization balance is enforced via
\begin{equation}
\frac{\left[ 1-\zeta(r)\right]^2}{\zeta(r)} = \frac{\Gamma(r)}{n(r)\alpha_{\rm B}}
\end{equation}
where $\alpha_B = 2.6\times 10^{-13}\,$cm$^{-3}\,$s$^{-1}$ at $T=10^4\,$K is the effective recombination coefficient for the on-the-spot approximation, and $n(r)$ is the total density. 

$\Gamma(r)$ is evaluated by the equation of radiative transfer. Recombinations to levels other than the ground state do not contribute photons for ionization, and photons from recombinations to the ground state have been accounted for in $\alpha_{\rm B}$. This simplifies the equation of radiative transfer substantially (no source term).
\begin{equation}
\Gamma(r) = 2\pi \int_0^\pi\, {\rm d}\theta \sin\theta \, J_{\nu_0}({\rm IGM})\,\exp^{-\tau(r,\theta)}
\end{equation}

where $\tau(r,\theta)$ is the optical depth to ionizing photons arriving at radius $r$ from the IGM, at a polar angle $\theta$:
\begin{equation}
\tau(r,\theta) = a_0\, \int_{0}^{\infty}\,{\rm d}x\, n(x)\zeta(x)\,f_{\rm v}(x)
\end{equation}
where the integrand is just the column density of neutral atoms along the ray, and $x$ is the affine parameter along the ray. The above equation will need to be evaluated numerically for a spherical geometry that we are considering here (unlike the plane-parallel approximation that is usually employed).

Taken together, we are now tasked with solving the following integral equation in $\zeta(r)$
\begin{equation}
\frac{\left[1-\zeta(r) \right]^2}{\zeta(r)}  = \frac{\Gamma({\rm IGM})}{2\alpha_{\rm B}n(r)}\,\int_0^{\pi}\,{\rm d}\theta\, \sin\theta \, \exp\left[-a_0\,\int_0^\infty\,{\rm d}x\,n(x)\zeta(x)f_{\rm v}(x) \right]
\end{equation}
Any given set of halo mass and redshift completely specify $n(x)$, $r_{\rm c}(x)$, $f_{\rm v}(x)$, and $\Gamma({\rm IGM})$. This allows us to solve for $\zeta(x)$ recursively via
\begin{equation}
\label{eqn:zeta_recursion}
\frac{\left[1-\zeta_{i+1}(r) \right]^2}{\zeta_{i+1}(r)}  = \frac{\Gamma({\rm IGM})}{2\alpha_{\rm B}n(r)}\,\int_0^{\pi}\,{\rm d}\theta\, \sin\theta \, \exp\left[-a_0\,\int_0^\infty\,{\rm d}x\,n(x)\zeta_i(x)f_{\rm v}(x) \right]
\end{equation}
We choose an initial value by setting the optical depth term to unity:
\begin{equation}
\frac{\left[1-\zeta_{0}(r) \right]^2}{\zeta_{0}(r)} = \frac{\Gamma({\rm IGM})}{\alpha_{\rm B}n(r)}
\end{equation}

\begin{figure*}
\centering
\includegraphics[width=\linewidth]{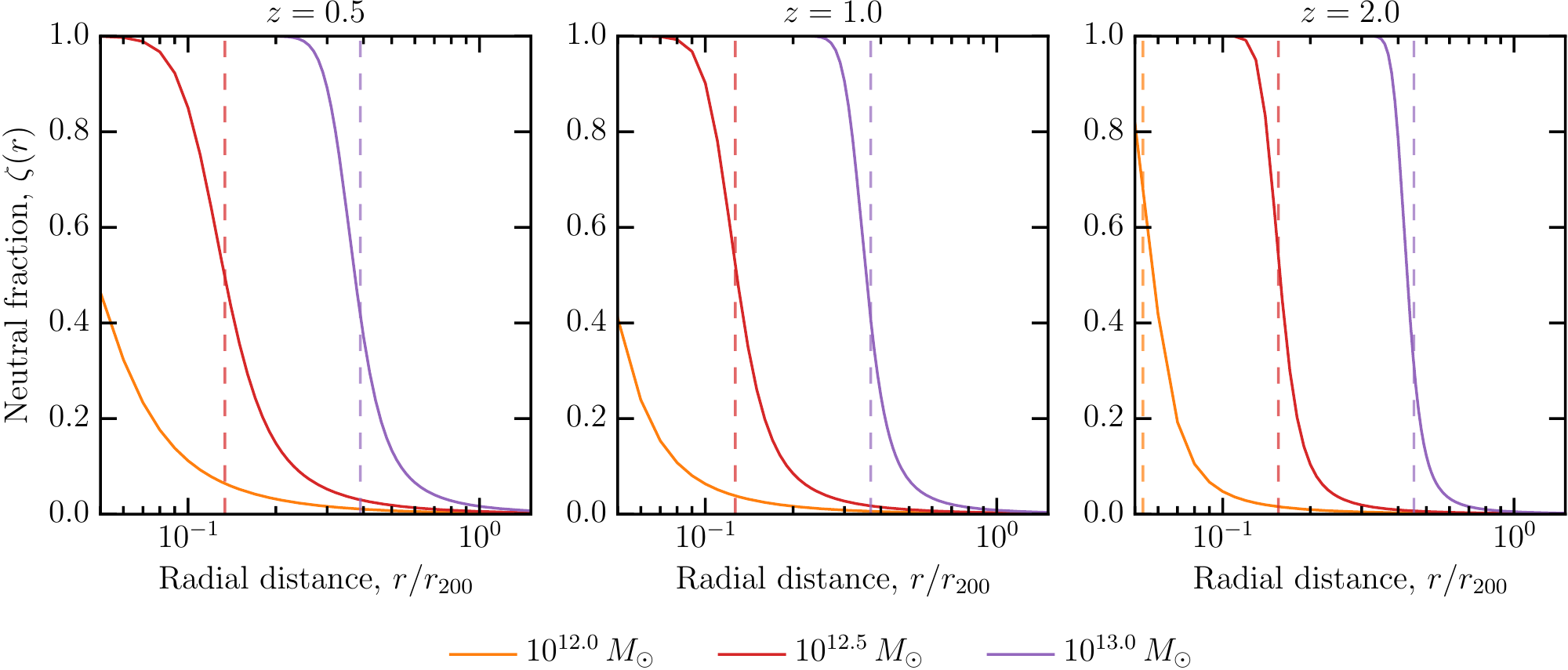}
\caption{Neutral fraction profile, $\zeta(r)$, computed using full radiative transfer of the IGM UV field from equation \ref{eqn:zeta_recursion} for different redshifts and halo mass (halo properties defined in \S 2). The dashed lines show the location of the ionization front as approximated by equation \ref{eqn:app_zeta_approx}.\label{fig:zeta_profile}}
\end{figure*}

An approximate location of the ionization front can be found by assuming (i) a radiation field given by $\Gamma({\rm IGM})$ throughout the halo to compute the neutral fraction in each cloud, and (ii) computing the radial depth at which the ensuing neutral fraction yields a column density of $a_0^{-1}$. Because individual clouds at the outskirts of the halo are only partially neutral, we have $\zeta(r) \approx n(r) \alpha_{\rm B}/\Gamma({\rm IGM})$. Setting the neutral column density integrated from the halo edge, $r_{\rm shock}=1.5\,r_{200}$ to the self-shielding radius $r_{\rm ss}$ in units of $r_{200}$, to $a_0^{-1}$, we get
\begin{equation}
a_0\alpha_{\rm B}/\Gamma({\rm IGM})r_{200}\,\int_{r_{\rm ss}}^{1.5}\,{\rm d}(r/r_{200})\, n^2(r/r_{200})f_{\rm v}(r/r_{200})=1
\end{equation}
We can let the upper limit of integration to recede to infinity without significant loss of accuracy, substitute for the radial scaling of density and volume fraction from \S 2, and get an approximate expression for the self-shielding radius in units of $r_{200}$
\begin{equation}
\label{eqn:app_zeta_approx}
r_{\rm ss} \approx 0.11\,\left(\frac{f_{\rm v}}{10^{-4}}\right)^{0.56}\,\left(\frac{f_{\rm CGM}}{0.3}\right)^{1.11}\,M_{12}^{0.93}\,\frac{h^{2.59}(z)}{(1+z)^{2.78}}.
\end{equation}
We have adjusted the numerical constant with a factor of order unity to match $r_{\rm ss}$ to the radius at which $\zeta=0.5$ (within $\sim 10$ per-cent) for mass and redshift ranges of interest, in the neutral profile determined from the full radiative transfer as per equation \ref{eqn:zeta_recursion}. Figure \ref{fig:zeta_profile} shows the neutral fraction profile evaluated using equation \ref{eqn:zeta_recursion} and the approximation for $r_{\rm ss}$ given above.

\label{lastpage}
\end{document}